\documentclass[onecolumn,12pt]{pasa}
\usepackage{graphicx}
\usepackage{hyperref}
\usepackage{tabularx}
\usepackage{ulem}
\usepackage[dvipsnames]{xcolor}

\usepackage{natbib}
\usepackage{amsmath}
\usepackage{verbatim}
\usepackage{gensymb}
\usepackage{textcomp}
\usepackage{MnSymbol}
\usepackage{aas_macros}
\usepackage{microtype,siunitx,booktabs}

\title{Understanding the impact of diffusion of CO in the astrochemical models \vspace{0.2cm}}
\author{Kinsuk Acharyya

\affil{Planetary Sciences Division, Physical Research Laboratory, Ahmedabad, 380009, India
\\ acharyya@prl.res.in}
}
\begin{document}
\maketitle
\begin{abstract}
The mobility of lighter species on the surface of interstellar dust grains plays a crucial role in forming simple through 
complex molecules. Carbon monoxide is one of the most abundant molecules, its surface diffusion on the grain surface is 
essential to forming many molecules. Recent laboratory experiments found a diverse range of diffusion barriers for CO  on 
the grain surface, their use can significantly impact the abundance of several molecules. The impact of different diffusion 
barriers of CO, in the astrochemical models, is studied to understand its effect on the abundance of solid CO and the species 
for which it is a reactant partner. A gas-grain network is used for three different physical conditions; cold-core and warm-up 
models with slow and fast heating rates. Two different ratios (0.3 and 0.5) between diffusion and desorption barrier are utilized 
for all the species. For each physical condition and ratio, six different models are run by varying diffusion barriers of CO. Solid 
CO abundance for the models with the lowest diffusion barrier yields less than 0.1 \% of water ice for cold clouds and a maximum of 
0.4 \% for slow and fast warm-up models. Also, solid CO$_2$ in dense clouds is significantly overproduced ($\sim$ 140 \% of water). 
The abundance of H$_2$CO and CH$_3$OH showed an opposite trend, and HCOOH, CH$_3$CHO, NH$_2$CO, and CH$_3$COCH$_3$ are produced in 
lower quantities for models with low diffusion barriers for CO. Considerable variation in abundance is observed between models with 
the high and low diffusion barrier. Models with higher diffusion barriers provide a relatively better agreement with the observed 
abundances when compared with the models having lower diffusion barriers.
\end{abstract}

\noindent{\it Keywords\/}:ISM, molecules, abundances, dust, extinction, astrochemistry
%\keywords{ISM, molecules, abundances, dust, extinction, astrochemistry}
\maketitle

\newpage
\section{Introduction}
In the cold ($\sim$ 10 K) and dense interstellar medium (ISM), solid carbon monoxide (CO) is ubiquitous. 
Its abundance varies from source to source but can be as high as 10$^{-4}$ relative to the total hydrogen 
density. It can form efficiently in the gas phase, which can be depleted on the surface of cold dust grains 
efficiently \citep{Caselli1999}. On the grain surface, CO can play a significant role in the formation of 
various molecular species \citep{Garrod2008}. Recent theoretical and laboratory measurements found a 
considerable variation in the CO diffusion barrier from one measurement to the other. In some cases, it is 
lower than that is commonly used in astrochemical models. Since surface diffusion is the key to the formation 
of molecules on the dust grain, therefore, it is important to re-visit the role of CO in the grain surface chemistry.

Molecules that are commonly found in the star-forming regions are partly or almost solely formed on the 
surface of the bare dust grains or on the water-rich ices that reside on the grains \citep{Herbst2008, Herbst2009}. 
Interstellar dust grains provide the surface for a reactant to meet with the other and take up the reaction's 
excess energy, which makes the reaction possible \citep{Hasegawa1992}. For reaction to occur on the grain surface, reactants need to 
have sufficient mobility so that they can scan the grain surface in search of a reaction partner. The mobility 
of an adsorbed species to migrate to an adjacent surface site can come from thermal hopping and quantum 
tunnelling. However, mobility due to thermal hopping is most commonly used in the astrochemical models. 
In the low temperatures ($\sim$ 10 K) of dense molecular clouds, only atomic hydrogen has sufficient 
mobility, making hydrogenation reactions the most important class of reactions on the surface of the grain. 
However, as the temperature rises due to the star formation process, other species like CO, N, O, OH, CH$_2$, 
CH$_3$, NH$_2$, NH$_3$, etc., become mobile and take part in the formation of complex molecules \citep{Garrod2008}.

%%%%%%%%%%%%%%%%%%%%%%%%
The hopping rate of any species on a dust grain is calculated through $t_{hop} = \nu exp(E_\textrm b/k_BT_d)$, where $E_\textrm b$ 
is the binding energy for thermal hopping, $T_d$ is the temperature of the dust grain and $\nu$ is the characteristic vibrational 
frequency for the adsorbed species \citep{Hasegawa1992}. In majority of astrochemical models the diffusion barrier, is assumed to 
be a fraction of desorption barrier ($E_\textrm d$), i.e., $E_\textrm b$ = $fE_\textrm d$. In most cases, a value of either 0.3 or 
0.5 is used for $f$ \citep{Hasegawa1992, Herbst2008, Herbst2009, Garrod2008, Wakelam2008}. The value of both the $E_\textrm d$ and 
$\nu$ can be estimated using temperature programmed desorption (TPD) experiments performed in the astrophysically relevant surfaces
\citep{Fraser2002, Collings2003, Bisschop2006, Noble2012, He2016a}. Then $E_\textrm b$ is estimated from $E_\textrm d$ by using suitable 
value for $f$. Direct measurement of diffusion barriers are difficult to make and sometimes comes with a large uncertainty. Some of the 
species for which diffusion barrier is measured either via theoretical calculations or experimental measurements include NH$_3$ 
\citep{Livingston2002, Mispelaer2013}, CH$_3$OH \citep{Livingston2002, Marchand2006}, atomic hydrogen \citep{Katz1999, Watanabe2010, 
Hama2012, Asgei2017, Sene2017}, atomic oxygen \citep{Minissale2013a, Minissale2013b, Minissale2014}, and CO \cite{Mispelaer2013, Lauck2015, 
Kars2014}.  Experimental determinations of the diffusion energy barrier for CO, N$_2$, O$_2$, Ar, CH$_4$ are given in \cite{He2018}.

The value of $E_\textrm b$ depends on several factors such as bare surface material (e.g., silicate, carbonaceous), nature of the 
binding sites, and surface roughness. Also, astrophysical grains at low temperatures ($\sim$ 10 K) are coated with layers of ices, 
which are water dominated. Therefore, often measurements are done with a layer of water ice on the top of the surface \citep{Hama2013}; 
ice layers of abundant molecules such as CO are also used \citep{Fraser2002, Collings2003, Bisschop2006}. 
Thus value of $E_\textrm b$ show a large variation due to change in substrate and its property.  For instance, the 
$E_\textrm b$ for hydrogen for olivine and carbonaceous surfaces are 24.7 and 44 meV, respectively \citep{Katz1999} and 
\cite{Watanabe2010} found two types of potential sites with the energy depths of $\sim$ 20 and $>$ 50 meV, respectively on 
water ice. Thus the hopping rate of hydrogen will be different for different use of $E_\textrm b$, which has a significant 
effect on hydrogenation reactions on the grain surface. Similarly, of diffusion barrier of CO \cite{Mispelaer2013, Lauck2015, 
Kars2014, He2018, Kouchi2020} is found to vary significantly from one experiment to the other. While hydrogenation reactions 
are relatively well studied, the effect of different CO diffusion barriers on astrochemical modelling has not been explored much. 

The main goal of this paper is to understand the impact of various diffusion barrier energies of CO, in the 
abundances of solid CO and surface species, which requires CO in their formation. In the next section, the barrier 
energy for CO-diffusion and desorption is discussed. In \S 3, the importance of solid CO in the surface 
chemistry, followed by the chemical networks and the model parameters in \S 4 and 5 are 
discussed. Results are presented in \S 6, the comparison with the observation in \S 7, and 
finally concluding remarks are made in \S 8.

%%% Checked is and are
\section{Barrier for CO-diffusion and desorption} \label{CO-Barrier}
%%%%%%%%%%%% Moved from introduction
In dense cold interstellar clouds, adsorption of CO on the dust grain is very efficient. 
%whether CO will take part in a reaction depends on its diffusion barrier ($E_\textrm b$) and the activation energy 
%barrier ($E_\textrm a$) of reaction, if any. 
Early astrochemical models, such as \cite{Allen1977} used $E_\textrm d$ 
for CO to be 2.4 kcal/mole ($\sim$ 1200 K), which makes  $E_\textrm b$ to be 360 and 600 K when $f$ = 0.3 and 0.5 
respectively. Subsequently, temperature programmed desorption (TPD) experiments are performed to study the desorption 
energy of CO  by several groups (\cite{Bisschop2006, Acharyya2007, Noble2012, Fayolle2016, He2016b}) on a variety of 
astrophysically relevant surfaces. These experiments found that $E_\textrm d$ for CO vary from as low as 831 $\pm$ 
40 K to as high as 1940 K, thus $E_\textrm b$ between 249 and 582 for $f$ = 0.3, and between 430.5 and 970 K when 
$f$ = 0.5. In addition; recent experiments found that desorption barriers are not only dependent on the nature of 
the surface but also coverage; $E_ \textrm d$ hence $E_\textrm b$ is lowest when coverage is about a monolayer or 
more and goes up with decreasing coverage \citep{He2016b}. The coverage dependence of $E_\textrm d$ and hence 
$E_\textrm b$ could be very important because, in interstellar clouds, the grain mantles are made of mixed ices, 
and layers of pure ices would not be possible for most species. A summary of various laboratory experiments to 
study $E_\textrm d$ for CO is listed in Table~\ref{TabB1}. If the barrier for diffusion is taken as a fraction of 
desorption barrier energy, it can vary significantly from one substrate to the other and on the coverage.

\begin{table*}
\centering
\caption{Desorption energies ($E_\textrm d$) for CO for selected experiments.}
\label{TabB1}
\begin{tabular}{|c|c|l|l|}
\hline
Substrate/ice          & $E_\textrm d$         & Comments                          & Reference           \\
                       & (K)                   &                                   &                     \\
\hline
Gold coated copper     & 855 $\pm$ 25          & multi-layered pure ice            & \cite{Bisschop2006} \\ 
                       & 930 $\pm$ 25          & multi-layered N$_2$-mixed ice     &                     \\ 
\cline{2-4}
                       & 858 $\pm$ 15          & multi-layered pure ice            & \cite{Acharyya2007} \\ 
                       & 955 $\pm$ 18          & multi-layered O$_2$-mixed ice     &                     \\ 
\hline
Crystalline water ice  & 849 $\pm$ 55          & multiyered                        &  \cite{Noble2012}   \\
\cline{2-3}
                       & 1330, 1288, 1199,     & when coverage = 0.1, 0.2,         &                     \\
                       & 1086 and 1009         & 0.5, 0.9 and 1 respectively       &                     \\
\cline{1-3}
Silicate surface       & 831 $\pm$ 40          & multilayered                      &                     \\
\cline{2-3}
                       & 1418, 1257, 1045,    & when coverage = 0.1, 0.2,          &                     \\
                       &    896 and 867       & 0.5, 0.9 and 1 respectively        &                     \\
\cline{1-3}
Nonporous water ice    & 828 $\pm$ 28          & multilayered                      &                     \\
\cline{2-3}
                       & 1307, 1247, 1135,     & for coverage = 0.1, 0.2, 0.5,     &                     \\
                       &  953 and 863          & 0.9 and 1 respectively            &                     \\
%\cline{1-3}
\hline
Non porous             & 870 - 1600$^1$        &  0.0056 - 1.33 ML                 & \cite{He2016b}      \\
amorphous water ice    &                       &                                   &                     \\
\cline{1-3}
Porous amorphous       & 980 - 1940$^1$        & 0.0056 - 1.33 ML                  &                     \\
water ice              &                       &                                   &                     \\
\cline{2-4}
                       & 1575 $\pm$ 117        & 0.7 ML                            & \cite{Fayolle2016}  \\
\cline{1-3}
Compact amorphous      & 866 $\pm$ 68          & multilayer ice                    &                     \\
water ice              & 1155 $\pm$ 133        & 1.3 ML                            &                     \\
                       & 1180 $\pm$ 131        & 0.8 ML                            &                     \\
                       & 1236 $\pm$ 139        & 0.3 ML                            &                     \\
                       & 1298 $\pm$ 116        & 0.2 ML                            &                     \\
\hline
Amorphous solid        & 1180                  & multilayer ice                    & \cite{Collings2003}  \\
\cline{2-4}
  water                & 1419                  & monolayer                         & \cite{Smith2016}    \\
\hline
CO$_2$ ice             & 1240 $\pm$ 90         &  multilayer ice                   & \cite{Cooke2018}     \\
                       & 1410 $\pm$ 70         &                                   &                      \\
\hline
\end{tabular}
\\
$^1$ Dependent on coverage.
\end{table*}

\begin{table*}
\centering
\caption{Diffusion energies ($E_\textrm b$) for CO for selected experiments.}
\begin{tabular}{|l|l|l|l|}
\hline
Substrate              & $E_\textrm b$ & Comments                          & Reference               \\
                       & (K)           &                                   &                         \\
\hline
Water and CO mixture   & 300 $\pm$ 100 & surface-segregation rate          & \cite{Oberg2009}        \\
Amorphous H$_2$O ice   & 158 $\pm$ 12  & Surface diffusion barrier$^{*1}$  & \cite{Lauck2015}        \\
Amorphous H$_2$O ice   & 120 $\pm$ 180 & Surface diffusion barrier$^{*2}$  & \cite{Mispelaer2013}    \\
H$_2$O ice             & 577 $\pm$ 12  & theoretical calculation           & \cite{Kars2012}         \\
H$_2$O ice             & 302 $\pm$ 174 & theoretical calculation           & \cite{Kars2014}         \\
H$_2$O ice             & 490 $\pm$ 12  & laboratory measurement            & \cite{He2018}           \\
H$_2$O ice             & 350 $\pm$ 50  & laboratory measurement            & \cite{Kouchi2020}       \\
                       &               & using UHV-TEM                     &                         \\
\hline
\end{tabular}
\\
${^*1}$ \cite{Lauck2015} found CO diffusion into the H$_2$O ice matrix is a pore-mediated process and described 
that the extracted energy barrier as effectively a surface diffusion barrier. \\
$^{*2}$ \cite{Mispelaer2013} obtained the value by fitting the experimental diffusion rates measured at different
temperatures with an Arrhenius law.
\label{TabB2}
\end{table*}

Experiments to estimate diffusion energy of CO directly was first attempted by  \cite{Oberg2009}, and found that 
surface-segregation rate follows the Arrhenius law with a barrier of 300 $\pm$ 100 K for CO on H$_2$O:CO mixture. 
One of the first experimental measurements to determine the activation energy for diffusion for CO from the porous 
amorphous ice is performed by \cite{Mispelaer2013}. They found a diffusion barrier of 1.0 $\pm$ 1.5 kJmol$^{-1}$ 
($\sim$ 120 $\pm$ 180 K) on porous amorphous ice, which makes E$_\textrm{CO, b}$/E$_\textrm{CO, d}$ $\sim$ 0.1. 
\cite{Kars2014} found that the CO mobility on the ice substrate is strongly dependent on the morphology, and it 
could have two sets of mobility, 30 meV ($\sim$ 350 K) for weakly bound sites and 80 meV ($\sim$ 930 K) for strongly 
bound sites. Subsequently, \cite{Lauck2015} using a set of experiments and applying Fick's diffusion equation to analyze 
the data, found the energy barrier for CO diffusion into amorphous water ice is 158 $\pm$ 12 K. Diffusion energy barriers 
for CO from various studies are listed in the Table~\ref{TabB2}. More recently, \cite{He2018} found a barrier of 
490 $\pm$ 12 K and \cite{Kouchi2020} found a barrier of 350 $\pm$ 50 K. Except for the theoretical calculation of 
\cite{Kars2012}, all the measurements found that the diffusion barrier is lower than the currently used values in 
the astrochemical models, which varies between 400 and 600 K depending upon E$_\textrm{b, CO}$/E$_\textrm{d, CO}$ 
used. Another important aspect is the value of $\nu$; all the measurements found a lower value with only exception 
is \cite{Kouchi2020} which used standard pre-exponential factor as mentioned by Equation~\ref{Eqn-2}. However, 
measurement of pre-exponential factor can have large uncertainty; its effect is discussed in \S \ref{pre-expo}.

\section{CO-surface Chemistry}\label{CO_Chem}
Solid CO is an important reactant on the surface of the grain and takes part in the formation of CO$_2$, CH$_3$OH, 
among many other species. The formation of CO$_2$ on grain surface can occur mainly via following reactions,
\begin{equation}
\rm{CO + H \rightarrow HCO \xrightarrow{\text{O}} CO_2 + H}
\label{eq:6}
\end{equation}
\begin{equation}
\rm{CO} + \rm{OH} \rightarrow \rm{CO}_2 + \rm{H}
\label{eq:7}
\end{equation}
\begin{equation}
\rm{CO + O \rightarrow CO_2.}
\label{eq:8}
\end{equation}
The first step in the Equation~\ref{eq:6} has a barrier between 4.1 and 5.1 kcal/mole (2110 - 2570 K) as 
calculated by \cite{Woon1996}. Lately, (\cite{Andersson2011}) found a barrier height of about 1500 K, which drops from the 
classical value due to tunnelling above a certain critical temperature. The activation energy of 2500 K is used for the calculation, 
although lowering the activation energy to 1500 K is also discussed. Similarly, Equation \ref{eq:7} to form CO$_2$ have a barrier 
which is found to go down from 2.3 kcal/mol (1160 K) to 0.2 kcal/mol (100 K) in the presence of water molecule (\cite{Tachikawa2016}).
Since interstellar ice is believed to be water rich, barrier of 100 K is the used for the calculations. The calculated barrier for 
Equation~\ref{eq:8}, varies from as low as 298 K (\cite{Roser2001}), to 1580 K (\cite{Goumans2008}). A barrier of 1000 K is used 
for the calculations which is average between these two values. However models with activation energy 
of 298 and 1580 K are also run and discussed. Formaldehyde and methanol, two very important molecules are formed via successive 
hydrogenation of CO as follows:
\begin{equation}
\rm{CO + H \rightarrow HCO \xrightarrow[\text{}]{\text {H}} H_2CO \xrightarrow[\text{}]{\text{H}} 
CH_3O \xrightarrow[\text{}]{\text{H}} CH_3OH}.
\end{equation}
These chain of reactions are extensively studied by several groups (\cite{Watanabe2003, Watanabe2004, Fuchs2009, Fedoseev2015, 
Song2017}). Similar to the first step, the third step also have a large barrier and it can form either CH$_2$OH or CH$_3$O, 
having a barrier of 5400 and 2200 K respectively (\cite{Ruaud2015}). Both the species can react with atomic hydrogen to form 
CH$_3$OH. Several reactions to form more complex molecules involving solid CO are considered following \cite{Garrod2008}. 
For example, Solid CO can form acetyl (CH$_3$CO) with an activation energy of 1500 K, and carboxyl group (COOH) as follows: 
\begin{equation}
\rm{CH_3 + CO \rightarrow CH_3CO},
\label{eq:9}
\end{equation}
\begin{equation}
\rm{CO + OH \rightarrow COOH}.
\label{eq:10}
\end{equation}
These groups can form more complex molecules by reacting with other species during cold core and warm-up phase. For example, 
hydrogenation reactions which are very efficient on the cold ($\sim 10 K$) surface of the grains can result in the formation 
of acetaldehyde from acetyl-group as follows:
\begin{equation}
\rm{ CH_3CO + H \rightarrow CH_3CHO,}
\end{equation}
and carboxyl group can form formic acid by the following reaction
\begin{equation}
\rm{H + COOH \rightarrow HCOOH}.
\end{equation}
Similarly, CO can react with NH and NH$_2$ to form NHCO and NH$_2$CO respectively (\cite{Garrod2008}). It also reacts with 
sulfur to make solid OCS on the grains. Also, HCO, CH$_3$O, COOH, NHCO, and NH$_2$CO can take part in the formation of several 
other complex organic molecules when grains are heated. Activation barrier used for various surface reactions involving CO is listed 
in Table~\ref{TabCO_React}. Thus diffusion rate of CO could have a significant impact in 
the formation of these molecules on the grain surface. Once produced on the grain surface, these molecules can come back to 
gas-phase via various desorption processes and can increase their respective gas-phase abundances.

\begin{table}
\centering
\caption{Activation barrier used for various surface reactions involving CO.}
\label{TabCO_React}
\begin{tabular}{|l|l|}
\hline
Reaction                              & $E_\textrm a$     \\
                                      & (K)               \\
\hline
CO + H        $\rightarrow$ HCO          & 2500$^a$       \\
HCO + O       $\rightarrow$ CO$_2$  + H  & 0$^b$          \\
CO + OH       $\rightarrow$ CO$_2$  + H  & 100$^c$         \\
CO + OH       $\rightarrow$ COOH         & 150$^d$        \\
CO + O        $\rightarrow$ CO$_2$       & 1000$^e$       \\
HCO + H       $\rightarrow$ H$_2$CO      & 0$^d$          \\
H$_2$CO + H   $\rightarrow$ CH$_3$O      & 2200$^d$       \\
H$_2$CO + H   $\rightarrow$ CH$_2$OH     & 5400$^d$       \\
CH$_3$ + CO   $\rightarrow$ CH$_3$CO     & 1500$^f$       \\
CH$_2$OH + CO $\rightarrow$ CH$_2$OHCO   & 1500$^f$       \\
CH$_3$O + CO  $\rightarrow$ CH$_3$OCO    & 1500$^f$       \\
NH + CO       $\rightarrow$ HNCO         & 1500$^f$       \\
NH$_2$ + CO   $\rightarrow$ NH$_2$CO     & 1500$^f$       \\
CO + S        $\rightarrow$ OCS          & 0$^f$          \\
\hline
\end{tabular}
\\
a. \cite{Woon1996}, b. \cite{Garrod2011}, c. \cite{Tachikawa2016}, d. \cite{Ruaud2015},  \\
e. The average value between \cite{Roser2001} and \cite{Goumans2008} is used, and \\
f. \cite{Garrod2008}.
\end{table}

%\begin{table}
%\centering
%\caption{Important model parameters are summarized.}
%\label{TabCO_React}
%\begin{tabular}{|l|l|l|l|l|l|}
%\hline
%Parameters                           & Ca                     & Cb           & M3         & M5         & M1                      \\
%\hline
%$E_\textrm{b, CO}$                   & $f(\theta$)$^{*1}$  & $f(\theta)$  & 360        & 600        & 158                     \\
%$E_\textrm{d, CO}$                   & $f(\theta$)         & $f(\theta$)  & 1200       & 1200       & 1200                    \\ 
%
%%$E_\textrm{b}$/$E_\textrm{d}$$^{*2}$ & 0.3                 & 0.5          & 0.3        & 0.5        & 0.3                     \\
%\hline
%\end{tabular}
%\\
%$^1$Desorption and diffusion energies are coverage dependent.
%$^2$Except for CO in M1, the ratio applies to all the species including CO.
%\end{table}

\begin{table}
\centering
\caption{Important model parameters are summarized.}
\label{Tab_Param}
\begin{tabular}{|l|ll|ll|ll|ll|ll|ll|}
\hline
Parameters         & \multicolumn{2}{c|}{C}          & \multicolumn{2}{c|}{M1} & \multicolumn{2}{c|}{M2} & \multicolumn{2}{c|}{M3} & \multicolumn{2}{c|}{M4} & \multicolumn{2}{c|}{M5} \\
\hline
                   & a                  & b          & a            & b        &       a    & b          &      a    & b           &      a    & b           &      a    & b           \\
\hline
$\mathcal{R}_{CO}$ & $f(\theta$)$^{*1}$ & $f(\theta$)& 0.1          & 0.1      &       0.2  & 0.2        &      0.3  & 0.3         &      0.4  & 0.4         &      0.5  & 0.5         \\
\hline
$\mathcal{R}_{i}$  & 0.3                & 0.5        & 0.3          & 0.5      &       0.3  & 0.5        &      0.3  & 0.5         &      0.3  & 0.5         &      0.3  & 0.5         \\
\hline
$E_\textrm{d, CO}$ & $f(\theta$)        & $f(\theta$) & 1150        &          &       1150 &            &     1150  &             &     1150  &             &     1150  &             \\
\hline
Symbol$^2$ & {\color{red} $\bullet$} & {\color{SkyBlue}$\filledmedtriangleup$} & {\color{black}$\times$}&& {$\color{GreenYellow}\filledstar$}&&{\color{blue}$\filledsquare$} &&
{\color{ForestGreen}$\filleddiamond$}& & {\color{magenta}$\square$}& \\
%$E_\textrm{b, CO}$    & $f(\theta$)$^{*1}$  & $f(\theta)$  & 360        & 600        & 158                     \\
%$E_\textrm{d, CO}$    & $f(\theta$)         & $f(\theta$)  & 1200       & 1200       & 1200                    \\ 

%$E_\textrm{b}$/$E_\textrm{d}$$^{*2}$ & 0.3                 & 0.5          & 0.3        & 0.5        & 0.3                     \\
\hline
\end{tabular}
\\
$^1$Desorption and diffusion energies are coverage dependent.
$^2$Same colour/symbols are always used with a combination line styles.
\end{table}

\section{Gas-Grain Chemical network}\label{Network}
To study the formation of molecules in the astrophysical conditions, one needs chemical networks that contain reaction rate 
constants. In the gas-phase, the rate constant ($\kappa_\textrm{i,j}(T)$) of bimolecular reactions in the modified Arrhenius 
formula is given by:
\begin{equation}
\kappa_\textrm {i,j}(T)=\alpha \left(\frac{T}{300} \right)^\beta exp \left(\frac{-\gamma}{T} \right) 
\textrm{cm}^3 \textrm{sec}^{-1},
\end{equation}
where, $\alpha$, $\beta$ and $\gamma$ are three parameters and $T$ is the temperature. Our gas-phase network is mainly based 
on  KIDA database (http://kida.obs.u-bordeaux1.fr), an acronym for Kinetic Database for Astrochemistry \citep{KIDA2015}. It 
includes reactions which are relevant for astrochemical modelling and similar to widely used databases such as OSU, UMIST. 
However, rate constants in the KIDA database come with four possible recommendations for users; not recommended, unknown, valid, and recommended.
The network also included ion-neutral reactions as described in \citet{Woon2009} and high-temperature reaction network from \cite{Harada2010}. 
Besides it update rates and new reactions regularly. In addition to this, assorted gas-phase reactions for complex molecules added 
from \cite{Garrod2008} and  additional reactions as described in \cite{Acharyya2015, Acharyya2017}.

The surface reaction rates are treated via rate constants in which diffusion is modelled by a series of hopping or tunnelling motions 
from one surface binding site to the nearest neighbour \citep{Hasegawa1992}. More exact treatments use stochastic approaches, however for a 
very large chemical network it's computationally very expensive. A two-phase model is considered in which the bulk of the ice mantle is not 
distinguished from the surface. In a three-phase model, surface and mantle are treated separately and considered chemically inert \citep{Garrod2011} 
or strongly bound compared to the surface hence less reactive\citep{Garrod2013}. The three-phase models allow the ice composition to be 
preserved through later stages for example, which allow carbon to be locked in CO, CO$_2$, CH$_4$ or CH$_3$OH but not as large hydrocarbons 
\cite{Garrod2011}. Thus a two-phase will tend to under-produce these ices and over-produce large hydrocarbons. However, the mobility of lighter 
species such as H, CO on the mantle is poorly understood and requires detail experimental studies to understand the true impact of three-phase 
models. Therefore a two-phase model is used for this study. There is no open database, which provides grain surface reactions and their rate 
constants. It can be calculated, analogous to those in the gas-phase two-body reactions by following (\cite{Hasegawa1992}). The first step 
is to calculate the thermal hopping rate ($r_\textrm{diff, i}$) of a species $i$ is given by
\begin{equation}
r_{{\textrm {diff, i}}} = \nu_0 exp(-E_\textrm {b, i}/T_\textrm d),
\label{Eqn-1}
\end{equation}
where, $T_\textrm d$ is the dust temperature, and $\nu_0$ the characteristic vibration frequency described by,
\begin{equation}
\nu_0 = \sqrt \frac{2n_sE_d}{\pi^2m},
\label{Eqn-2}
\end{equation}
where, n$_s$ is the surface density of sites ($\sim$ 1.5 $\times$ 10$^{15}$ cm$^{-2}$) and m is the mass of the 
absorbed particle.

Then the surface reaction rate $k{_\textrm{ij}}^{'}$ (cm$^3$s$^{-1}$) is given by
\begin{equation}
R_\textrm{ij} = k{_\textrm{ij}}^{'}n_\textrm s(i)n_\textrm s(j),
\end{equation}
where n$_\textrm s$(i), is the surface concentration (N$_\textrm i$.n$_\textrm d$) of species i and the rate coefficient 
$k{_\textrm{ij}}^{'}$ 
is given by 
\begin{equation}
k{_\textrm{ij}}^{'}=\kappa_\textrm{ij}(r_\textrm{diff, i} + r_\textrm{diff, j})/n_\textrm d.
\end{equation}
The parameter $k_\textrm{ij}$ is unity when there is no barrier for exothermic reactions. However, when there is an activation 
energy barrier ($E_\textrm A$), the rate is modified by multiplying a simple factor, which is either a tunnelling probability 
($k_{ij, t}$) or a hopping probability, whichever is greater. The factor due to tunnelling probability is given by
\begin{equation}
k_{ij, t} = \nu exp[-2(a/\hbar)(2\mu E_A)],
\end{equation}
where $a$ is the width of the potential barrier and $\mu$ is the reduced mass of the i-j system (\cite{Hasegawa1992}). The 
hopping probability is given by multiplying the rate with $exp(-E_A/T)$. Both the hopping and tunnelling probability are
compared, and whichever is greater is used for the calculation. 

The grain-surface network 
used for this study has three major components: (i) a network involving complex organic species from \cite{Garrod2008}, 
(ii) non-thermal desorption mechanisms including photodesorption processes by external UV photons and cosmic-ray generated UV photons, 
and (iii) additional reactions discussed in \cite{Acharyya2017}.

\section{Model parameters} \label{Model_param}
Models are run by varying $E_\textrm{b, CO}$/$E_\textrm{d, CO}$,  $E_\textrm{b, i}$/$E_\textrm{d, i}$, and physical conditions. 
In total 36 models are run.  For easier description, let us define $E_\textrm{b, CO}$/$E_\textrm{d, CO}$ as $\mathcal{R_{CO}}$ 
and $E_\textrm{b, i}$/$E_\textrm{d, i}$ as $\mathcal{R}_i$. Models are run for six different sets of CO diffusion barriers, 
which are designated as M1, M2, M3, M4, M5, and C. First we varied the $\mathcal{R_{CO}}$ between 0.1 and 0.5, where 
$E_\textrm{d,CO}$ = 1150 K, which is presently used in the astrochemical models. The ratio 0.1 (Model M1) is representative 
of very low diffusion barrier measured by \cite{Lauck2015, Mispelaer2013, Kars2014}. In model M3 and M5, $\mathcal{R_{CO}}$ = 
0.3 and 0.5 respectively are considered, which are most commonly used ratio's in the models. Thus CO diffusion energy for M3 and 
M5 is 345 and 575 K respectively. Besides, $E_\textrm{b, CO}$/$E_\textrm{d, CO}$ = 0.3 is close to measured value of 350 $\pm$ 50 K 
by \cite{Kouchi2020}. In models M2 and M4 $\mathcal{R_{CO}}$ = 0.2 and 0.4 are used respectively. For all the five values of 
$\mathcal{R_{CO}}$, we ran models with $\mathcal{R}_i$ = 0.3 and 0.5, which are designated as a and b, e.g., for model M1a, 
$E_\textrm{d, CO}$ = 0.1 and $\mathcal{R}_i$ = 0.3. Similarly, for M1b, $E_\textrm{d,CO}$ = 0.1 but $\mathcal{R}_i$ = 0.5 is 
used. Then coverage dependent binding energies are used with $\mathcal{R_{CO}}$ and $\mathcal{R}_i$ = 0.3 (Ca) and 0.5 (Cb) 
following \cite{He2016b}. The equation for coverage dependent binding energy for desorption is given by,
\begin{equation}
E_d(\theta) = E_1 + E_2 exp \left( -\frac{a}{max(b - log(\theta), 0.001)} \right)
\end{equation}
where E$_1$, E$_2$, a, and b are fitting parameters. E$_1$ is the binding energy for $\theta$ $\ge$ 1 Mono Layer (ML), 
which is 870 K for CO and,  while [E$_1$ + E$_2$ (730 K)] is the binding energy when $\theta$ approaches zero. Thus E$_d$ 
varies between 870 and 1600 K, therefore the barrier energy for diffusion varies between 261 and 480 K for Ca 
($\mathcal{R}_{CO} = 0.3$) model and between 435 and 800 K for Cb ($\mathcal{R}_{CO} \sim 0.5$) model. To find 
coverage of a species its abundance is divided with the abundance of one monolayer. It makes total 12 sets of parameters, 
which are listed in Table~\ref{Tab_Param}.

All the 12 models are run for three different physical conditions. The first one represents cold cores for which all the 
physical conditions remain homogeneous with n$_\textrm H$ = 2 $\times$ 10$^4$ cm$^{-3}$, gas and dust temperature = 10 K, 
and visual extinction ($A_\textrm V$) = 10 mag. For other two models, the two-phase physical model as prescribed by 
\cite{Brown1988} is followed. In the first phase or the free fall collapse phase, the cloud undergoes isothermal collapse at 
10 K, from a density of 3000 to 10$^7$ cm$^{-3}$ in about 10$^6$ yrs. During which visual extinction grows from 1.64 to 432 mag. 
In the second phase, collapse is halted, and the temperature is increased linearly from 10 to 200 K over two time-scales; 
5 $\times$ 10$^4$ and 10$^6$ yrs which are representative heating rates for low and high mass star formation respectively 
(\cite{Garrod2008}). Finally, chemical evolution of the hot core phase is continued till the total time evolution reaches 10$^7$ yrs.

The lifetime for the hot core is larger considering the high density of the hot core phase. However, the total time of 10$^7$ 
years is chosen in analogy with the dense cloud models for the sake of completeness. In the plots, warm-up regions and a short 
period after that is zoomed. It is important to note that the nature of free fall collapse phase is such that the increase of 
density and visual extinction is very slow for most of the times. Therefore timescales for the formation of molecules will be 
larger, compared to the standard dense cloud models with a fixed density of 2 $\times$ 10$^4$ cm$^{-3}$ at early time. 
However, density and visual extinction rise very rapidly towards the very end of free fall collapse phase, making timescales for 
formation smaller. It is expected that this aspect will be reflected in the 
time evolution of various species.

The sticking coefficient is close to unity for the isothermal models kept at 10 K, whereas, for warm-up models, temperature-dependent 
sticking coefficient is used following \cite{He2016a}. We included reactive desorption with $a_\textrm{RRK}$ value set at 0.01, 
i.e., about 1 \% of the product will be released to the gas-phase upon formation via grain-surface chemistry. The effect of reactive 
desorption  at early times (t $\le$ 2 Myr) on the gas-phase abundance profile is almost negligible except for few hydrogenated 
species \citep{Garrod2007}. It also does not alter the peak abundances. However, at the late time, it reduces the extent of depletion 
and helps to maintain gas-phase chemistry by re-injecting the various species from the grain surface \citep{Garrod2007}. Cosmic ray 
ionization rate $\zeta$(s$^{-1}$) of 1.33 $\times$ 10$^{-17}$ is used for all the models. All grains were assumed to have the same 
size of 0.1 $\mu$m, the dust-to-gas mass ratio of 0.01, and so-called low-metal elemental abundances are used (\cite{Wakelam2008}).

\section{Results}\label{Results}
Emphasis is given on grain surface abundances, particularly, CO and CO$_2$ ices. Gas phase abundances are discussed briefly. 
%The gas-phase abundances of molecules like CO, which are efficiently formed in the 
%gas-phase, are not expected to get directly affected due to the diffusion rates on the grain surface. 
In cold cores and the collapse phase of the warm-up models, dust temperature is $\sim$ 10 K, which allows CO to stick to 
the grain surface. It can take part in a surface reaction or desorb back to the gas-phase due to the thermal and various 
non-thermal desorption processes, which 
can cause a change in their gas-phase abundances. Discussion of the simulation results of six models corresponding to 
the cold cores is done with more details. Grain surface reactions involving CO are effective up to $\sim$ 20 - 25 K, 
above which it desorbs back to the gas-phase. In Table~\ref{TabOBS}, the observed abundances and peak model abundances are 
provided with for comparison. All abundances are shown with respect to the total hydrogen density unless mentioned otherwise.
All comparisons between model and observed abundances are for ices. 
Five models were run by varying $\mathcal{R}_{CO}$ ($E_\textrm{b, CO}$/$E_\textrm{d, CO}$) between 0.1 and 0.5 and 
designated as M1, M2, M3, M4, and M5, and one model is run with coverage-dependent binding energy for CO designated using `C'. 
Each such model is further classified using the alphabet `a' and `b' with $E_\textrm{b, i}$/$E_\textrm{d, i}$ = 0.3 and 0.5 
respectively. Symbols and line-styles for various models are kept same for all the Figures unless mention otherwise.
The circle (red), triangle (cyan), $\times$ (black),
asterisk (yellow), square (blue), diamond (green), and empty square (magenta) markers represent abundances for Model Ca, Cb, 
M1, M2, M3, M4, and M5 respectively for all the plots unless mentioned otherwise. Different line styles are used to distinguish between
`a' and `b' models. 
\begin{figure} 
\centering
\includegraphics[width=\columnwidth]{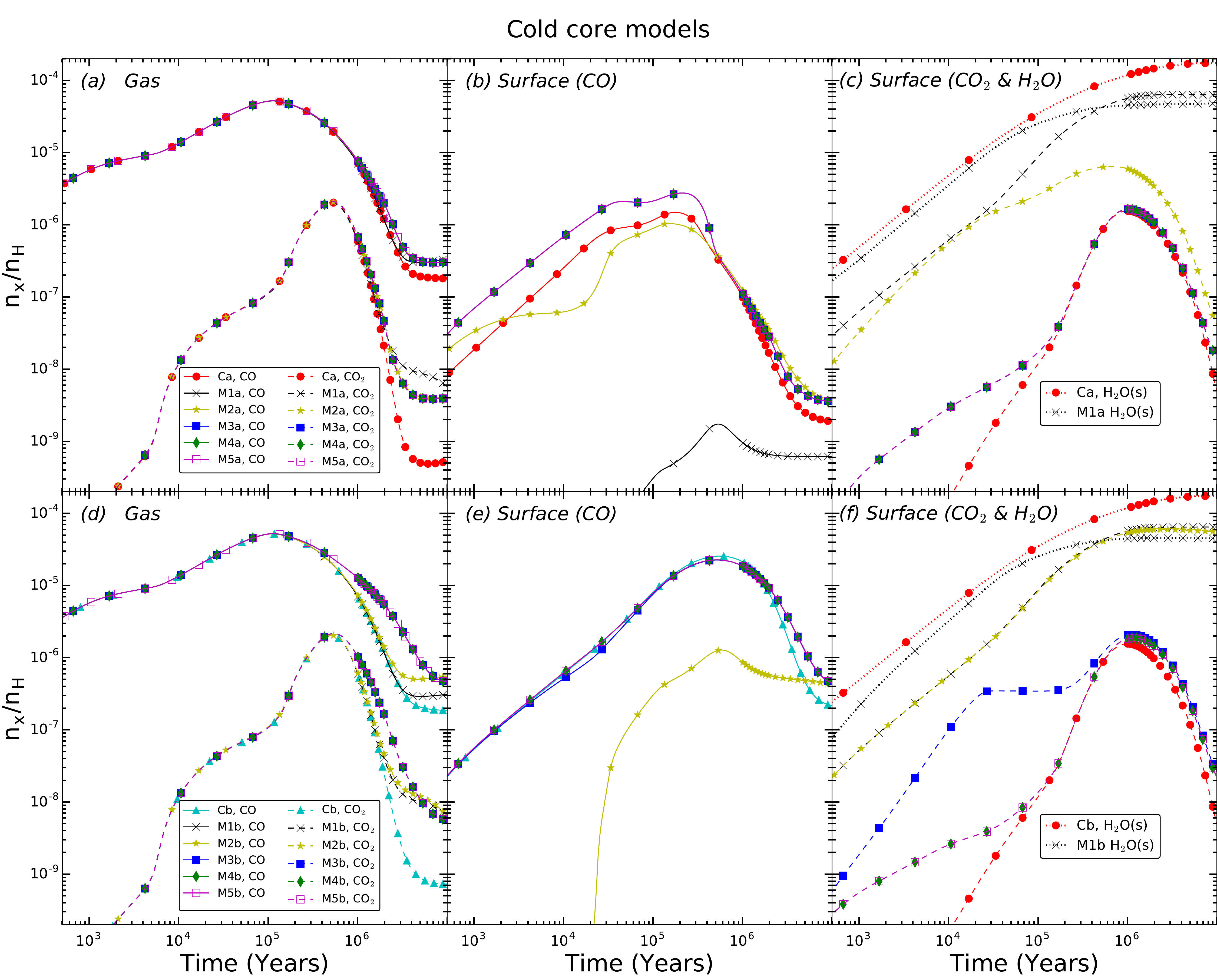}
\caption{
Time variation of the abundance of CO (solid lines) and CO$_2$ (dashed lines) for cold-core models are shown. Surface water 
abundance for Ca/Cb and M1a/M1b models are shown for reference using dotted lines. Panel (a) shows gas-phase abundances of CO 
and CO$_2$, (b) and (c) show surface abundances of CO and CO$_2$ respectively. All the abundances are plotted with respect to 
the total hydrogen density. The top panel is for $\mathcal{R}_i$ = 0.3 and bottom panel for 0.5.
%Circle (Red), triangle (cyan), $\times$ (black), asterisk (yellow),  square (blue), diamond (green), and empty square (magenta) represent abundances 
%for Model Ca, Cb, M1, M2, M3, M5, and M5 respectively. 
Legends for CO and CO$_2$ are same for all the panels. For colour figures 
please see the online version.
}
\label{fig_CO}
\end{figure}

\subsection{\rm{CO and CO$_2$}} 
Figure~\ref{fig_CO} shows the time variation of the gas-phase abundance of CO and CO$_2$ for all the six models 
that belong to the cold cores. The top and bottom panels show abundances for $\mathcal{R}_i$ = 0.3 
and 0.5 respectively, for different $\mathcal{R}_{CO}$.
The solid and dashed lines represent CO and CO$_2$ abundances, respectively. It is evident that gas-phase CO abundance profile 
(Figure~\ref{fig_CO}a, d) for all the models is almost same till 5 $\times$ 10$^5$ years, and 
peak abundance comes around at 5.2 $\times$ 10$^{-5}$ and at $\sim$ 10$^5$ yrs. Deviation in CO abundances for different 
models is not large and stays within a factor of 2.5 between the maximum (model M5) and minimum value (model Ca for 
$\mathcal{R}_i$ = 0.3 and Cb for 0.5). 
The gas-phase abundance for CO$_2$ is also mostly similar for all the six models, and it only deviates at the late 
times like CO. The pattern of deviation is also similar to that of CO, i.e., the model M5 has the maximum abundance 
and model Ca/Cb the least. The difference in abundance between these two models is nearly one order of magnitude. 
Interestingly, for the most of time evolution, the abundance of CO and CO$_2$ for model M1 for which CO diffusion 
is fastest is closer to the model abundance of M5, for which diffusion rate of CO is slowest. Thus gas-phase CO and 
CO$_2$ abundances got affected only at late times. It is expected since CO is efficiently formed in the gas-phase, 
the only way by which a difference could occur is via desorption processes. In cold cores, thermal 
desorption is ineffective, and the effect of non-thermal desorption is seen at late times. Therefore, deviations 
in gas-phase abundances are seen at late times.

In the Figure~\ref{fig_CO}b and Figure~\ref{fig_CO}e, CO abundances for models with $\mathcal{R}_i$ 
($E_\textrm{b, i}$/$E_\textrm{d, i}$) = 0.3 and 0.5 are plotted respectively for varying $\mathcal{R}_{CO}$ ($E_\textrm{b, CO}$/$E_\textrm{d, CO}$).
The abundance of surface H$_2$O for Ca/Cb and M1a/M1b models is also shown in the Figure~\ref{fig_CO}c/f using the dotted lines 
for reference. In the top panel (Figure~\ref{fig_CO}b) with $\mathcal{R}_i$ = 0.3, the profiles for $\mathcal{R}_{CO}$ = 0.3 (M3), 
0.4 (M4), and 0.5 (M5) have almost no difference with a peak value of, 2.8 $\times$ 10$^{-6}$ (5 \% of water), which is followed by 
Ca (1.5 $\times$ 10$^{-6}$, 3 \% of water), M2a (1 $\times$ 10$^{-6}$, 2 \% of water), 
M1a (1.7 $\times$ 10$^{-9}$, $<$ 0.01 \% of water). In the bottom panel ($\mathcal{R}_{i}$ = 0.5), models Cb, M3b, M4b, 
and M5b have almost no difference and having a peak abundance between 2.3 - 2.6 $\times$ 10$^{-5}$ (27 - 30 \% when compared with water). 
Model M2b ($\mathcal{R}_{CO}$ = 0.2) have a peak abundance of 1.3 $\times$ 10$^{-6}$ (3 \% of water), whereas M1a ($\mathcal{R}_{CO}$ = 0.1) 
model has a very low abundance of solid CO. Thus it is clear the solid CO abundances increases when its mobility decreases but once 
the $\mathcal{R}_{CO} \ge 0.3$, the change in abundance is small. The solid CO abundance is very low for models M1a, 
b and M2a, b due to the fast recombination of CO to form other species.  Also except for Model M1, solid CO abundance 
is greater for $\mathcal{R}_{i}$ = 0.5 (bottom panel) when compared between $\mathcal{R}_{i}$ = 0.3 (top panel), due 
to the slower conversion of CO into the other species.

\begin{figure}
\centering 
\includegraphics[width=8cm]{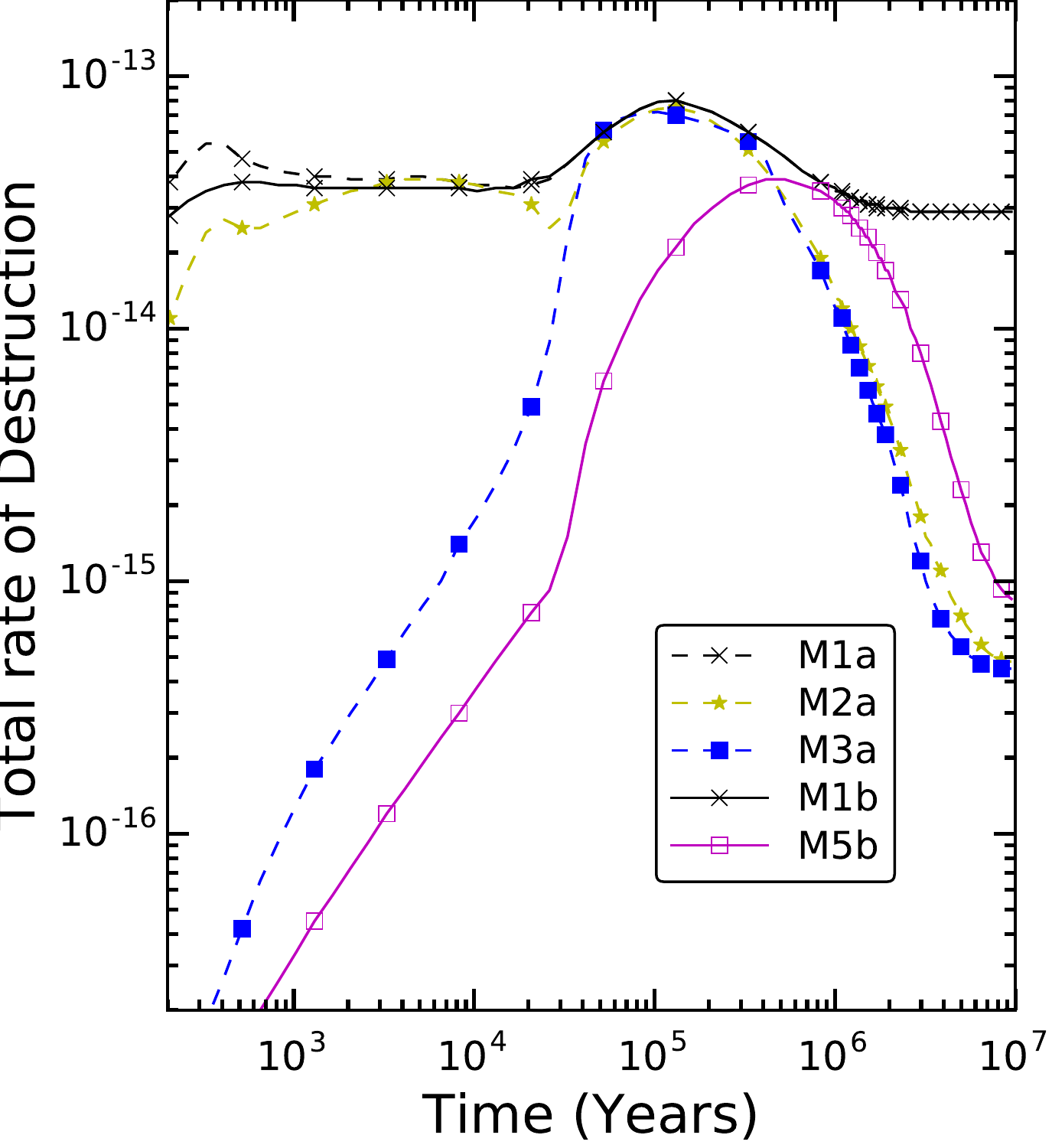}
  \caption{
Time variation of the total rate of destruction (cm$^{-3}$s$^{-1}$) of solid CO is shown for assorted cold-core models. For colour figures 
please see the online version.
}
\label{fig_CO_rate}
\end{figure}

To investigate farther, in Figure~\ref{fig_CO_rate}, the total destruction rate (r$_\textrm{CO, pd}$) which is obtained by 
adding all the individual rates in which CO is a reaction partner is plotted as a function of time for the assorted models.
It is evident that the r$_\textrm{CO, pd}$ for model M1 ($\mathcal{R}_{CO}$ = 0.1) and M2 ($\mathcal{R}_{CO}$ = 0.2)
is always higher compared to other models, especially at smaller $t$, when it is significantly higher than the other models. 
For other models r$_\textrm{CO, pd}$, gradually increases with time. 
The solid CO is primarily destroyed by the following two reactions:
\begin{equation}
CO + H \rightarrow HCO,
\label{desby_H}
\end{equation}
and
\begin{equation}
CO + OH \rightarrow CO_2 + H.
\label{desby_OH}
\end{equation}
For both ($E_\textrm {b, i}$/$E_\textrm {d, i}$ = 0.3 and 0.5) versions of C, M3, M4, and M5 the solid CO is primarily destroyed 
by Eq.~\ref{desby_H}, i.e., via hydrogenation. 
%For model M2, at very early time %(between 50 and 400 yrs), surface CO is destroyed by Eq.~\ref{desby_OH}, after that by hydrogenation. 
Whereas, for Model M1 and M2, the most dominant destruction pathway is always Eq.~\ref{desby_OH}. Thus when CO mobility is large 
($\mathcal{R}_{CO} \le 0.2$), formation of CO$_2$ is preferred, otherwise it is hydrogenated to form HCO.

The peak total destruction rate for both the versions of C, M1, M2, M3, M4 is $\sim$ 8 $\times$ 10$^{-14}$  cm$^{-3}$s$^{-1}$ and 
comes between 10$^5$ and 2 $\times$ 10$^5$ yrs. For models, Cb and M5b, the peak r$_\textrm{CO, pd}$ is lower by a factor of two and 
comes at a later time (around 4 $\times$ 10$^5$ yrs). It implies that due to higher diffusion barrier, CO is destroyed relatively 
slowly. Thus solid CO abundance is higher for these models.

Time variation of abundance of solid CO$_2$ is shown for all the models in the Figure~\ref{fig_CO}c and f for cold cores. Except 
for models M1a and M2a, for all the other models, the peak solid CO$_2$ abundance varies between (2 - 2.2) $\times$ 10$^{-6}$ 
(between 1 - 2 \% of water). For model M1a ($\mathcal{R}_{CO}$ = 0.1 and $\mathcal{R}_{i}$ = 0.3), CO$_2$ is produced very 
efficiently as shown in Figure~\ref{fig_CO}c. Peak solid CO$_2$ reaches close to the water ice abundance, which is the most 
abundant ice on the dust grains. Also, from Figure~\ref{fig_CO}c, it is clear that for the model M1 water abundance is lower 
by at least a factor of two when compared to other models. Since both CO$_2$ and water formation requires OH radicals, an efficient 
CO$_2$ formation due to faster CO diffusion rate, reduces OH radicals available for water formation, thereby decreasing its abundance.
Besides for models M1a and M2a the abundance of CO$_2$ starts to build-up earlier in time and remains flat. In the bottom panel, 
abundance for $\mathcal{R}_{i}$ = 0.5 is shown. The abundances are similar to as $\mathcal{R}_{i}$ = 0.3, except for M2b model 
($\mathcal{R}_{CO}$ = 0.2), for which the solid CO$_2$ abundance is higher compared to the M1b. 

\begin{figure}
\centering
\includegraphics[width=\columnwidth]{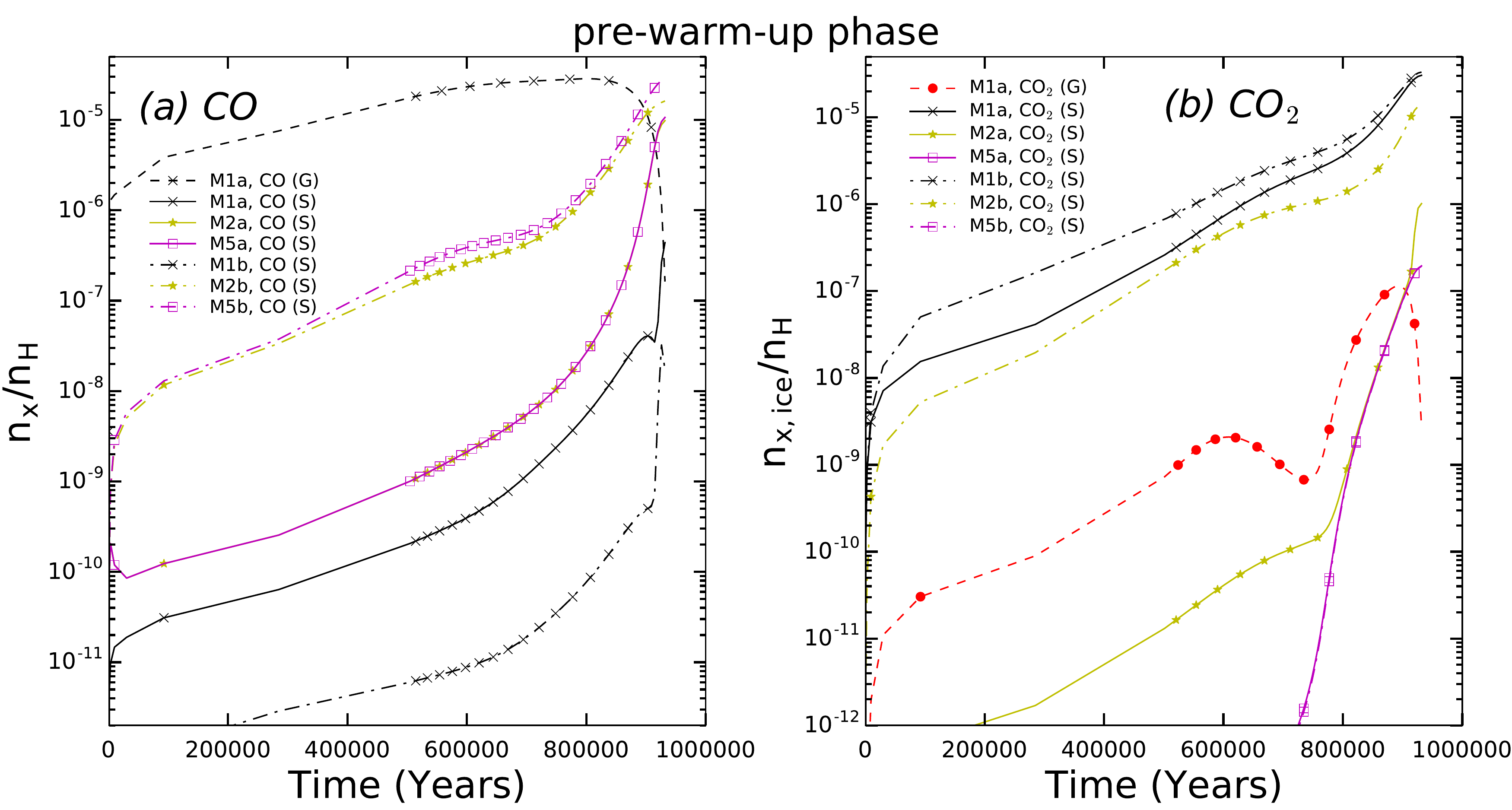}
  \caption{Time variation of CO (a) and CO$_2$ (b) for assorted models are shown for the pre-warm-up phase. Gas phase models with dashed lines are 
shown for M1 model only, since the abundance is similar. For colour figures please see the online version.
%The surface abundance of CO for M1 through M5 model is shown, Ca is similar to M3 and Cb is similar to M5. For CO$_2$ Ca, Cb, M4 and M5 is not 
%shown since they are similar to M3 model.
}
\label{fig_CO_collapse-Pre}
\end{figure}

\begin{figure} 
\includegraphics[width=\columnwidth]{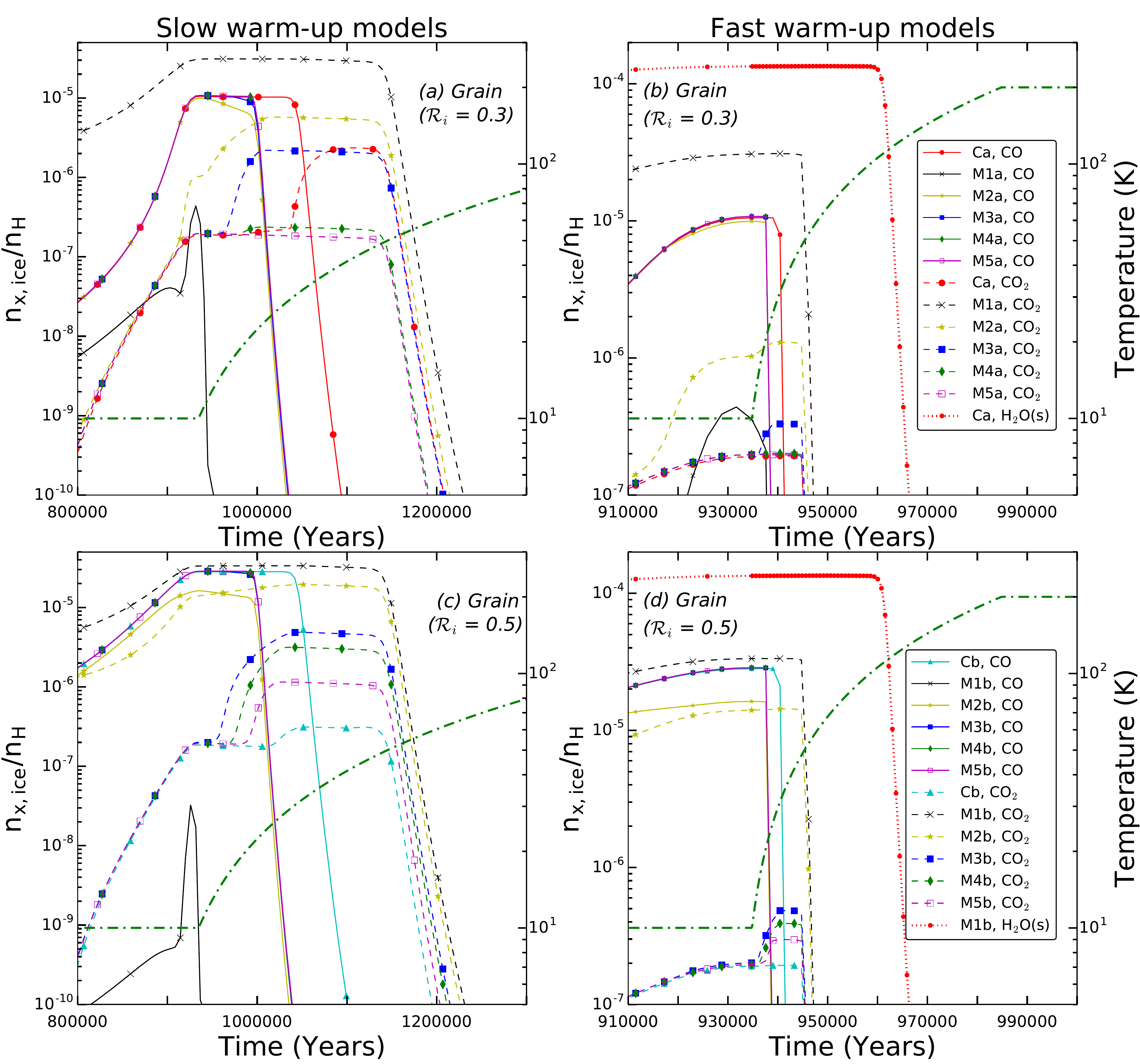}
  \caption{
Time variation of the abundance of CO (solid lines), CO$_2$ (dashed lines), are shown for all the warm-up models for 
$\mathcal{R}_{i}$ = 0.3 and 0.5 in top and bottom panel respectively. The surface water (dotted line) for Ca model is shown 
for reference (other models have similar abundances). Panels (a) and (c) are for low mass star formation (slow heating), 
whereas (b) and (d) are for high mass star formation (fast heating). 
%Circle/triangle, square, cross, star, square, diamond, and 
%empty square, markers represent abundances for Model Ca/Cb, M1, M2, M3, M4, and M5 models respectively for all the plots. 
Legends for CO and CO$_2$ applies for all the panels. Gas/grain temperature as a function of time is shown using dot-dashed 
(green) line and scale is provided in the opposite y-axis. For colour figures please see the online version. 
}
\label{fig_CO_collapse-a}
\end{figure}

%Figure~\ref{fig_CO_collapse-a} shows the time variation of the abundance of gas-phase and surface abundance of CO and CO$_2$ for 
%the warm-up models. It also shows water abundance for Ca model for reference (Figure~\ref{fig_CO_collapse-a}d). The abundance variation 
%for assorted models during the pre warm-up phase is shown in Figure~\ref{fig_CO_collapse-Pre}, in which density and visual extinction 
%are changing while temperature is constant as described in the \S 4.

Abundance profiles of CO and CO$_2$ for warm-up models are shown in the Figure~\ref{fig_CO_collapse-Pre} and 
\ref{fig_CO_collapse-a}. As described in the \S 5, that the warm-up models have two phases. The abundance 
profile of the first phase in which density and visual extinction are changing at constant temperature (10 K) is shown 
in the Figure~\ref{fig_CO_collapse-Pre}. In the pre warm-up phase, the surface CO abundance increases with decreasing 
diffusion rate, whereas for CO$_2$ show the opposite trend, that is the model M1 which have fastest diffusion have the 
highest abundance. Also difference in gas-phase abundances of CO and CO$_2$ between various models during the pre-warm-up 
phase is small. 

Figure~\ref{fig_CO_collapse-a} shows the abundances of solid CO for the warm-up and the post-warm-up phases 
for $\mathcal{R}_i = 0.3$ and $0.5$ in the top and bottom panel, respectively. For both $\mathcal{R}_i =$ 0.3 
and 0.5, models with $\mathcal{R}_{CO}$ = 0.1 (M1a and M1b) have the lowest abundance ($\sim$ 0.4 \% of water). 
Thus faster diffusion of solid CO resulted in the formation of CO$_2$. For other models, solid CO abundance is 
always large for $\mathcal{R}_i =$ 0.5 with a peak value of $\sim$ 3 $\times$ 10$^{-5}$ ($\sim$ 22 \% of water) 
when compared with $\mathcal{R}_i =$  0.3, which show a peak abundance of $\sim$ 1 $\times$ 10$^{-5}$ ($\sim$ 
8 \% of water). The model abundances are similar for both the fast and slow warm-up models. The models (Ca and Cb) 
for which coverage dependent barrier energies are used, CO desorbed at relatively higher temperatures when compared 
to the other models since its barrier for desorption increases with decreasing coverage. Since by $\sim$ 20 K, 
almost the entire CO is back to the gas phase due to the onset of thermal desorption due to warm-up, involvement 
of solid CO in the grain surface chemistry above this temperature will be limited.

The surface CO$_2$ abundances for the slow warm-up models are shown in the Figure~\ref{fig_CO_collapse-a}a 
($\mathcal{R}_{i} = 0.3$), and c ($\mathcal{R}_{CO} = 0.5$). In both cases solid CO$_2$ abundances increase as 
$\mathcal{R}_{CO}$ is decreased. The peak solid CO$_2$ abundance for both $\mathcal{R}_{i} = 0.3$ and $0.5$ comes 
for M1a and M1b model, which is around $\sim$ 3 $\times$ 10$^{-5}$ ($\sim$ 30 \%). There is a notable difference 
between M2a and M2b between $\mathcal{R}_{i} = 0.3$ and $0.5$, for the later case solid CO$_2$ is higher by a factor 
of four. It can be due to the slower diffusion of other species which resulted lesser consumption of solid CO. For 
faster warm-up similar behaviour for M1 and M2 model is observed, but for other models there is almost no difference 
in the solid CO$_2$ abundance profile.

For both CO and CO$_2$, the surface abundance is similar for models with $\mathcal{R}_{CO}$ 
($E_\textrm{b, CO}$/$E_\textrm{d, CO}) > $ 0.2 for all the three physical conditions. Thus above a certain critical 
diffusion barrier, abundances of CO of CO$_2$ species remain unchanged due to lack of CO mobility. 

A large difference in the solid CO$_2$ abundance in the M1 model (CO diffusion is the fastest) between warm-up 
and dense cloud models are seen. In the dense cloud, CO$_2$ abundance ($\sim$ 140\% of water) for M1 model becomes 
greater than water abundance but only a moderate increase ($\sim$ 30\% of water) is observed in the corresponding 
warm-up models. It is clear from Figure~\ref{fig_CO_collapse-Pre}a gaseous CO in the warm-up models increases rapidly 
at late times. Therefore, CO accreted to the grain surface towards the end of the free fall collapse phase (density 
rises much faster towards the end). However by that time a significant amount of O/OH required to form CO$_2$ ice is 
used up to form other species especially water; which resulted in the lower abundance of solid CO$_2$ for the M1 model 
in warm-up models. On the other hand, in the dense cloud, the density and visual extinction are always relatively higher 
at the early phase, which resulted in depletion of more CO on the grain relatively earlier than warm-up models. Thus 
physical conditions played a role to negate the effect of faster diffusion in the warm-up models.

\subsection{\rm{HCO, H$_2$CO, CH$_3$OH, and OCS}}
\begin{figure}[h] 
\includegraphics[width=\columnwidth]{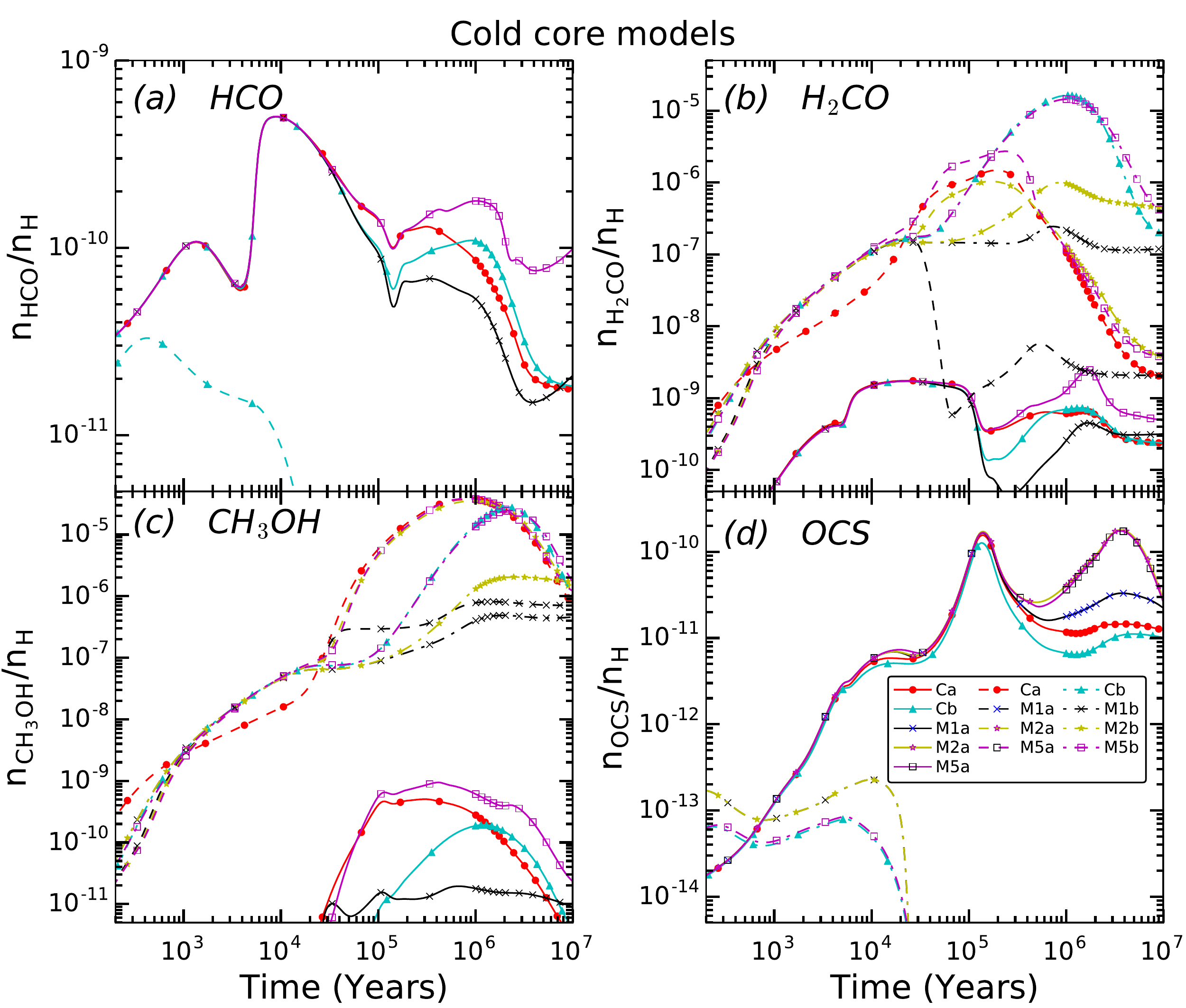}
  \caption{
Time variation of gas (solid lines) and surface (dashed lines) abundance of HCO, H$_2$CO, CH$_3$OH, and OCS for the assorted cold core models 
are shown. Model abundances of M3 and M4 are similar to model M5, therefore not shown. For colour figures please see the online version.
}
\label{fig_Oth_small}
\end{figure}

Figure~\ref{fig_Oth_small} show the time variation of the abundance of HCO, H$_2$CO, CH$_3$OH, and OCS for the assorted 
cold core models. The solid and dashed lines represent the gas-phase and grain surface abundances respectively. The results 
demonstrate that surface abundance of HCO and OCS are always low, whereas surface abundance is high for H$_2$CO and CH$_3$OH. 
Also, models with $\mathcal{R}_{CO}$ ($E_\textrm{b, CO}$/$E_\textrm{d, CO}) > 0.2$, have almost similar surface abundances 
for all the species.

For cold core models, the peak gas-phase abundance of HCO is almost same for all the models although abundance profile starts 
to diverge for time $>$ 4 $\times$ 10$^4$ years. The divergence is mainly due to various types of non-thermal desorption such as 
reactive desorption, due to which $\sim$ 1 \% is desorbed into the gas-phase. On the surface, HCO is efficiently produced, but 
abundance is low for all the models due to very efficient hydrogenation to form H$_2$CO, which have a large abundance. When HCO 
is produced, a small fraction is also ejected to the gas-phase, which can cause a difference in the gas-phase HCO abundance, provided 
its destruction rate in the gas-phase is slower than the rate of reactive desorption. Besides, Formation and destruction of HCO, H$_2$CO, 
and CH$_3$OH are closely linked, for an example the most 
dominant gas-phase formation route at late times for HCO is due to the reaction between H$_2$COH$^+$ and electron. The production 
of H$_2$COH$^+$ in the gas-phase occurs via the reaction between H$_2$CO/CH$_3$OH with various ions, e.g., H$_3$$^+$. Thus reactive 
desorption of H$_2$CO/CH$_3$OH will also play a role in the abundance variation of gas-phase HCO. 
%In the literature, reactive 
%desorption is considered as non-thermal desorption (\cite{Garrod2007}). Therefore it is counter-intuitive to find that diffusion 
%barrier is indirectly related to the non-thermal desorption. 
The solid HCO has a somewhat higher abundance for models Cb and M5b with $\mathcal{R}_{i}$ = 0.5. 

%For the warm-up models (Figure~\ref{fig_Oth_small_WU_LM}), gas-phase abundance of 
%HCO is higher compared to the cold core models, although solid HCO is still very low.

\begin{figure} 
\centering
\includegraphics[width=8 cm]{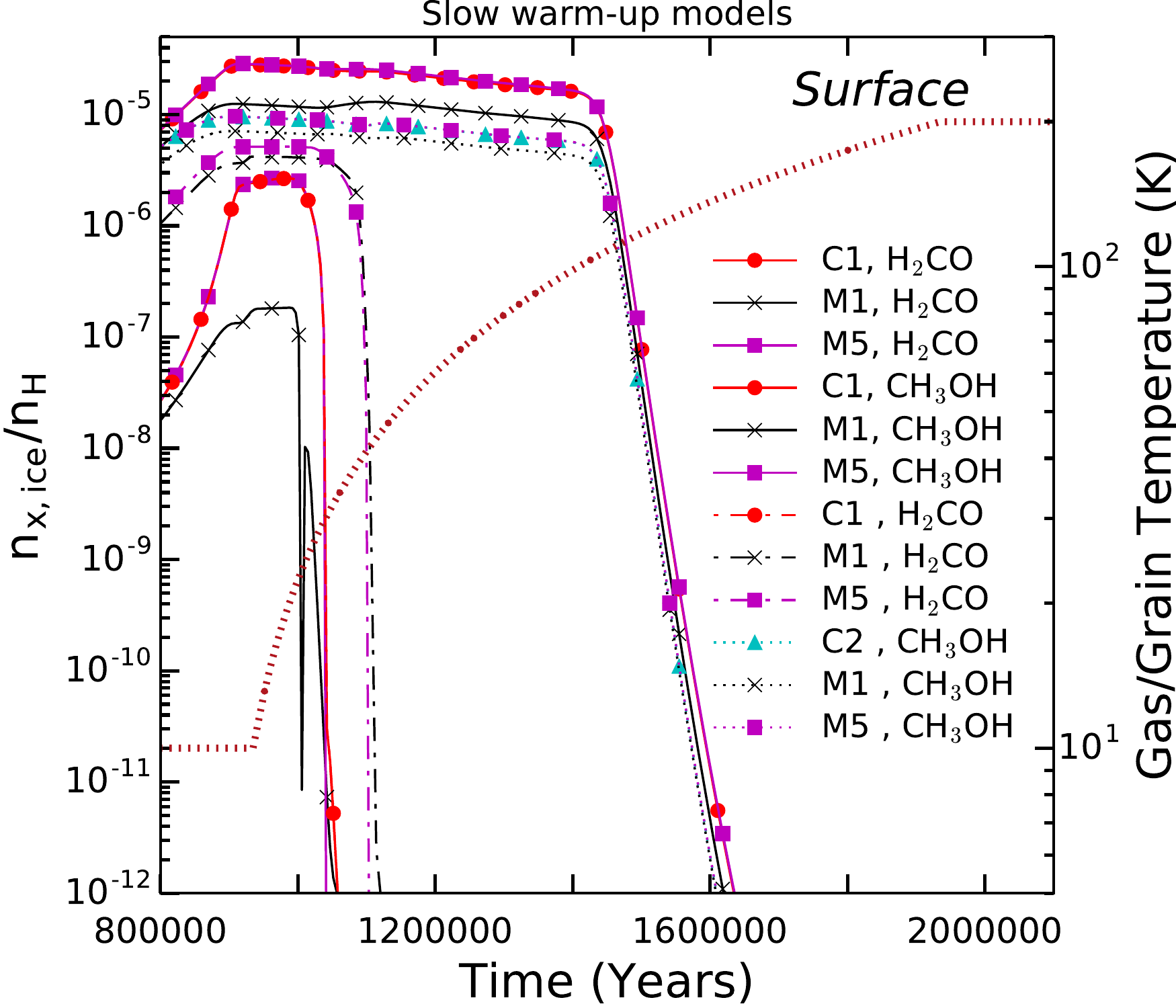}
  \caption{
Time variation of surface abundance of H$_2$CO and CH$_3$OH, for assorted models with slow warm-up are shown. 
Gas/grain temperature as a function of time is shown using dotted line (red) and scale is provided in the opposite y-axis. 
Legends apply to both the plots.}

\label{fig_Oth_small_WU_LM}
\end{figure}

Figure~\ref{fig_Oth_small}b show the abundance variation of H$_2$CO for assorted cold core models. Although gas-phase abundance is 
similar for all the models except, with a moderate variation ($\sim$ factor of three) in abundance between different models after 
10$^5$ yrs, the surface H$_2$CO varies significantly from one model to the other. Highest peak abundance for surface H$_2$CO is 
1.6 $\times$ 10$^{-5}$ (15.6 \% of water) for Cb, followed by the M5b model for which abundance is 1.5 $\times$ 10$^{-5}$ (14 \% 
of water). For other models, peak surface abundance is reduced by at least a factor of six and model M1a having the lowest 
abundance. It is to be noted that despite the high surface abundance, gas-phase abundance of H$_2$CO is low, due to the absence of 
efficient desorption mechanisms at 10 K.

Figure~\ref{fig_Oth_small}c shows the abundance variation of CH$_3$OH for assorted dense cloud models. It is primarily 
formed on the surface of the dust grains. Abundance profile of CH$_3$OH also exhibits that for $\mathcal{R}_{CO} > 0.2$, 
variation in abundance from one model to the other is small, which is true for both $\mathcal{R}_i = $ 0.3 and 0.5. Models 
with $\mathcal{R}_i = $ 0.3 have higher surface abundance compared to the corresponding models with $\mathcal{R}_i = $ 0.5.
Maximum solid CH$_3$OH is formed for $\mathcal{R}_i = $ 0.3 and $\mathcal{R}_{CO} \ge 0.2$, which is $\sim$ 3.5 $\times$ 10$^{-5}$ 
($\sim$ 38 \% of water). For models with $\mathcal{R}_i = $ 0.5 and $\mathcal{R}_{CO} > 0.2$ peak abundance is 2.5 $\times$ 10$^{-5}$ 
(20 \% of water). The lowest abundance of surface CH$_3$OH comes for the M1b model. Also, peak abundances for models with 
$\mathcal{R}_i = $ 0.5 comes at later times when compared with models for $\mathcal{R}_i = $ 0.3. 
It can be seen that for H$_2$CO, peak abundance comes for $\mathcal{R}_i = $ 0.5 models, whereas for CH$_3$OH, it comes for the 
$\mathcal{R}_i = $ 0.3. It is due to the slow conversion of H$_2$CO to CH$_3$OH due to slower diffusion. 
Whereas, for CH$_3$OH, models with $\mathcal{R}_i = $ 0.3 had a higher abundance due to faster diffusion, which results in 
quick conversion of H$_2$CO to CH$_3$OH via hydrogenation. For both the species, model M1 for which CO diffusion is the fastest, 
the abundance is the lowest since CO quickly converted to CO$_2$ instead 
of getting hydrogenated. 

Finally, Figure~\ref{fig_Oth_small}d shows the abundance variation of OCS for assorted cold core models; all 
the models have nearly the same peak gas-phase abundance and similar profiles till about 10$^5$ years. The abundance varies 
only at the later times. The gas-phase abundance of OCS is low $\sim$ 10$^{-10}$ to 10$^{-11}$; therefore, a small change 
in the abundance of more abundant reactants which are involve in its formation can account for the late time variation. The OCS 
in the gas-phase is mostly produced either from HOCS$^+$ or SO, which reacts with ions like HCO$^+$, CH$_3$$^+$, H$_3$$^+$, etc. 
to produce OCS. Besides, a major complication comes because OCS can also be converted to HOCS$^+$, so their production is interlinked. 
Solid OCS abundance is low for all the cold core models.

Selected slow heating warm-up model abundances for solid H$_2$CO and CH$_3$OH are shown in 
Figure~\ref{fig_Oth_small_WU_LM}; abundance does not vary from one model to the other except model for M1, for which abundance is lower 
by a factor of 10. The time variation of abundance of CH$_3$OH also show similar trends.
Abundance of HCO and OCS are small $< 10^{-10}$ therefore not shown.

\begin{figure} 
\includegraphics[width=\columnwidth]{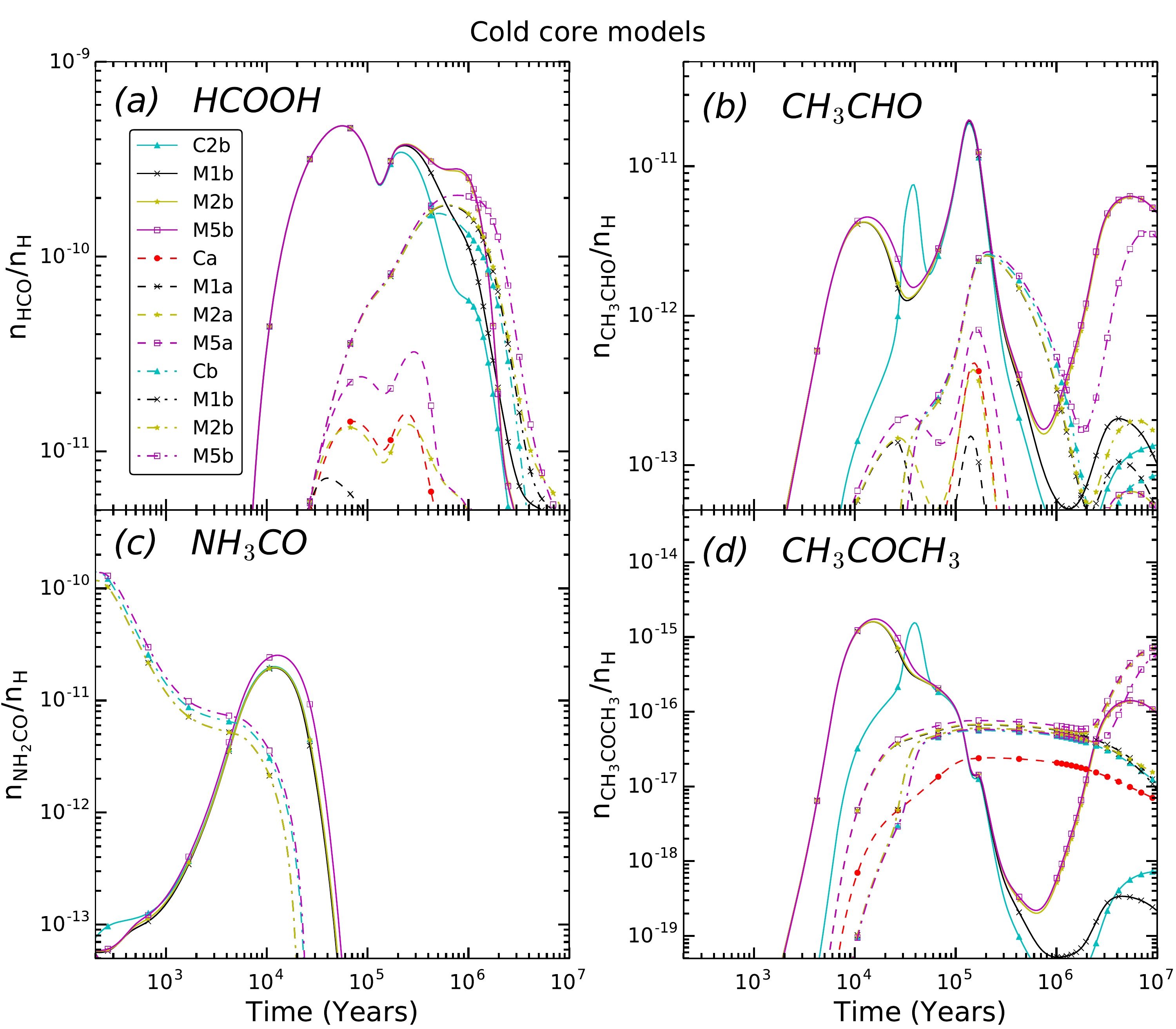}
  \caption{
Time variation of gas (solid lines) and surface (dashed lines for $\mathcal{R}_i = 0.3$, and dot-dashed lines for $\mathcal{R}_i = 0.5$) 
abundance of  HCOOH, CH$_3$CHO, NH$_3$CO, and CH$_3$COCH$_3$ for assorted cold core models are shown. Legends apply to all. 
For colour figures please see the online version.
}
\label{fig_Oth_big}
\end{figure}

\subsection{\rm{HCOOH, CH$_3$CHO, NH$_3$CO, and CH$_3$COCH$_3$}}
Time variation of gas (solid lines) and surface (dashed lines for $\mathcal{R}_i = 0.3$, and dot-dashed lines for $\mathcal{R}_i = 0.5$) 
abundance of HCOOH, CH$_3$CHO, NH$_3$CO, and CH$_3$COCH$_3$ for the assorted dense cloud models are shown in the Figure~\ref{fig_Oth_big}. 
Since surface abundances of these species are low, warm-up 
model results are not discussed although peak abundances are shown in Table~\ref{TabOBS}. Only species in the bunch which could have a 
noticeable amount of surface abundance is HCOOH. Its, gas-phase abundance is similar for all the models except the usual deviation at 
late times, while the surface abundance of HCOOH is lower and varies from one model to the other after 10$^4$ yrs. The models for which 
$\mathcal{R}_i = 0.5$, have nearly one order of magnitude higher abundance compared to the models having $\mathcal{R}_i = 0.3$. Also,
 for both the cases abundances are nearly same for $\mathcal{R}_{CO} > 0.2$.
Compared to the dense cloud and fast warm-up models, HCOOH abundance increases for slow warm-up models. 
For acetaldehyde (CH$_3$CHO), the gas-phase 
abundance for all the models is similar till about 10$^6$ yrs and having similar peak abundances, except for Cb model. Its surface abundance varies 
significantly from one model to the other, although the abundance is low; therefore, it will be difficult to detect. 
Both the formamide (NH$_3$CO) and acetone (CH$_3$COCH$_3$) have low gas-phase and surface abundance for all the models. 
Thus although there are variations in abundances for HCOOH, CH$_3$CHO, NH$_3$CO, and CH$_3$COCH$_3$ from one model 
to the other; none of the models increase the production significantly so that their abundance is above the present day 
detection limit on the ice. The capability of detection of ice is expected to increase significantly due to JWST 
mission, which can conclusively detect molecules like HCOOH. Also, formation mechanism of these complex molecules need to 
be revisited since at present they are not efficient enough.

\begin{figure}
\centering 
\includegraphics[width=\columnwidth]{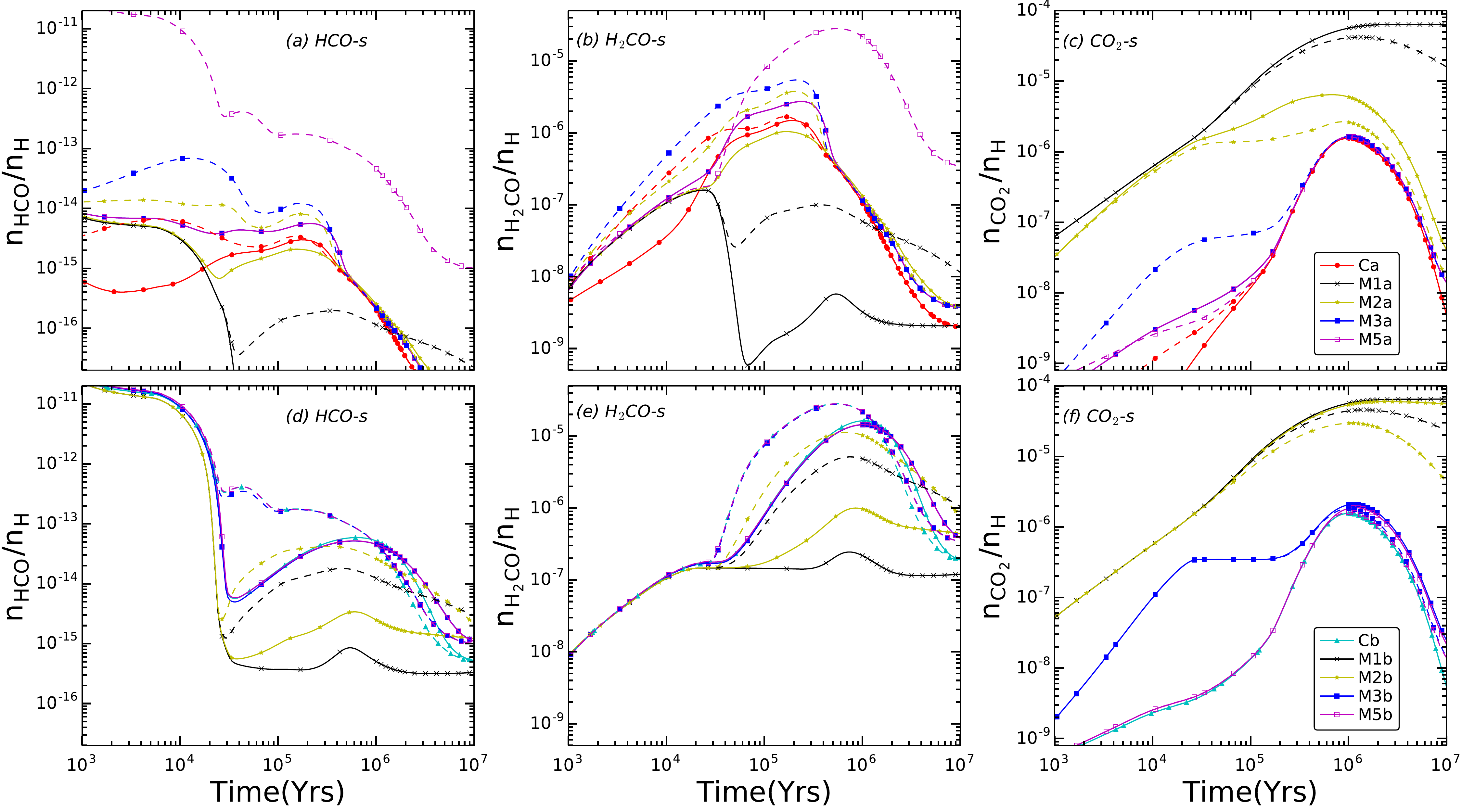}
  \caption{
Time variation of abundance of assorted species when activation energy for H + CO reaction is reduced to 1500 K (dashed lines) following 
\cite{Andersson2011} from 2500 K (solid lines) for the cold core models. For colour figures please see the online version.
}
\label{fig_Acti}
\end{figure}

\subsection{Effect of Activation Energy}
It is pertinent to discuss the effect of activation energy for the reactions for which multiple measurements are available. 
For the first step in Reaction~\ref{eq:6}, we have presented results having the activation energy of 2500 K. In an another
measurement \cite{Andersson2011} found the activation energy to be as low as 1500 K. Figure~\ref{fig_Acti} compares 
abundances between these two activation energies for cold core models. Models with activation energy of 1500 K shows higher 
HCO abundance for all the models for both $\mathcal{R}_{i}$ = 0.3 and 0.5. 
HCO can form H$_2$CO upon hydrogenation or CO$_2$ by reacting with atomic oxygen. Since hydrogen is much more mobile than oxygen, 
H$_2$CO is produced in higher quantity for all the models as can be seen in Figure~\ref{fig_Acti}b. For CO$_2$ the trend is reversed 
for models with $\mathcal{R}_{CO}$ = 0.1 and 0.2, for other models deviation is small.
For CO$_2$ formation via \ref{eq:8}, 
different activation energies (298 and 1580 K) did not yield any significant difference since it is not the most dominant formation 
pathways for CO$_2$ formation for any model.

\subsection{Effect of pre-exponential factor}\label{pre-expo}
\begin{figure*} 
\centering
\includegraphics[width=\columnwidth]{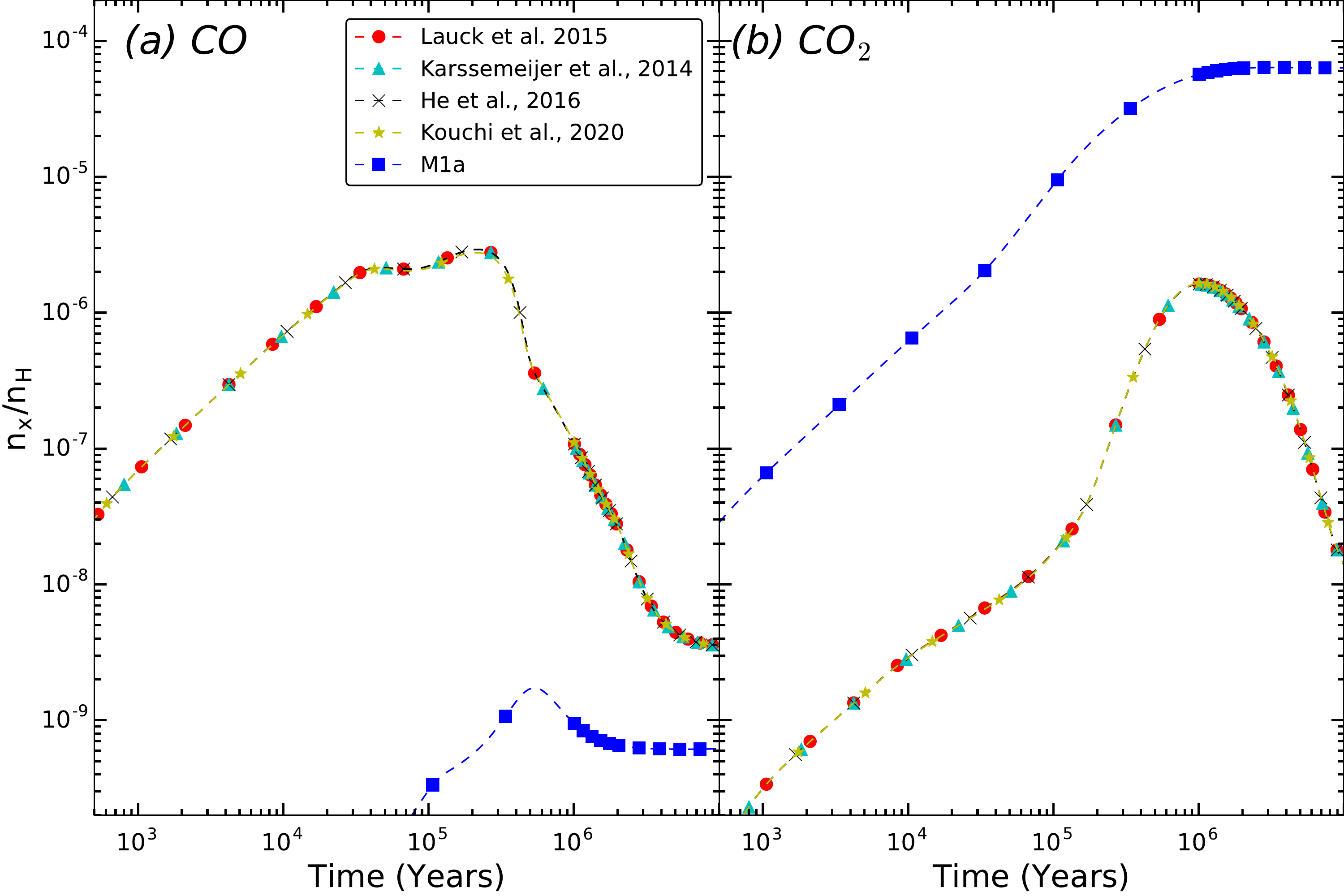}
\caption{
Time variation of solid abundance of CO and CO$_2$ with various combination of E$_b$ and $\nu_0$ and model M1a.
Legends applies to both the panel.
}
\label{fig_Com-Nu}
\end{figure*}

The diffusion rate is also depends on pre-exponential factor ($\nu_0$) as evident from the Equation~\ref{Eqn-1}, therefore, 
it is pertinent to discuss its effect. The solid CO measurements of \cite{Kars2014} and \cite{Lauck2015} reported D$_0$ 
values of 9.2 $\times$ 10$^{-10}$ and 3.1 $\times$ 10$^{-12}$ cm$^2$ s$^{-1}$ respectively. The value of $\nu_0$ can be 
roughly estimated by assuming, $\nu_0 = D_0/a^2$, where, $a \sim 3 \times 10^{-8}$cm is the typical hopping distance, which 
comes out to be $\sim$ 10$^{6}$ and $\sim$ 3 $\times$ 10$^{3}$ s$^{-1}$ for \cite{Kars2014} and \cite{Lauck2015} respectively. 
Also, \cite{He2018} found this value in the range of 10$^8$ - 10$^9$ s$^{-1}$. Where as \cite{Kouchi2020}, considered  
pre-exponential factor using Equation~\ref{Eqn-2} for their model. To test the dependence on $\nu_0$, four additional models 
are run with E$_b$ from \cite{Lauck2015},  \cite{Kars2014}, \cite{He2018}, and  \cite{Kouchi2020} with $\nu_0$ of 3 $\times$ 10$^{3}$, 
$\sim$ 10$^{6}$,10$^8$, and 10$^{12}$ s$^{-1}$ respectively. The time variation of abundance of CO and CO$_2$ for these models 
along with the M1a model are shown in the Figure~\ref{fig_Com-Nu}. It can clearly be found that there is almost no difference 
between these models. Abundances for these models are similar to M3a model abundance and significantly different than model M1a. 
Outcome is in line with other models, i.e., for $\mathcal{R}_{CO} > 0.2$, i.e, for slower diffusion rates of CO no appreciable 
difference in abundance is found for both CO and CO$_2$.

\section{Comparison with Observations}\label{Com_Obs}
\begin{figure*} 
\centering
\includegraphics[width=14 cm]{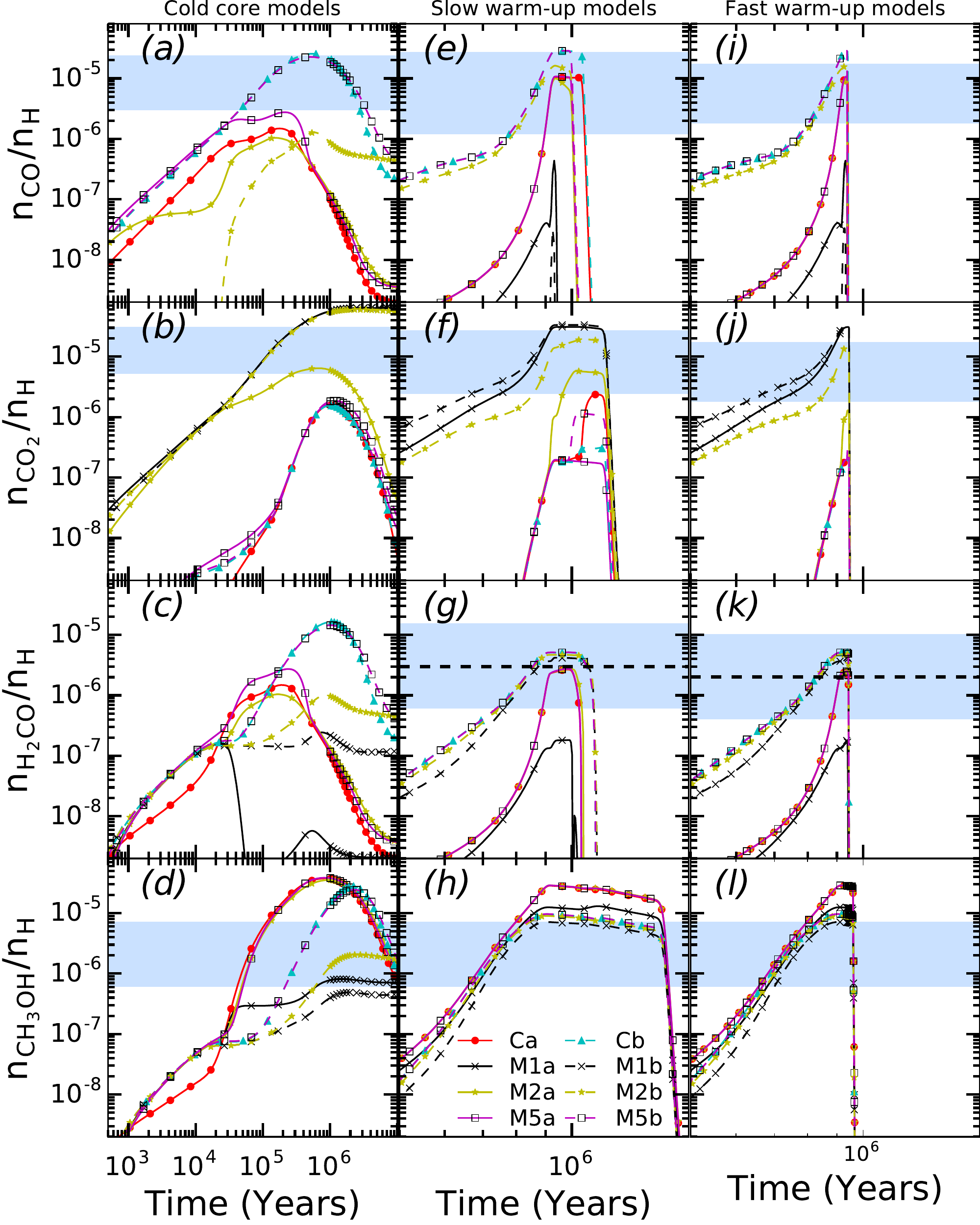}
\caption{
Comparison of observed abundances of CO, CO$_2$, H$_2$CO, and CH$_3$OH are shown for cold cloud (a, b, c ,and d), slow (e, f, g, and h) 
and fast (i, j, k, and l) heating warm-up models. Shaded regions represent the observed range for solid CO, CO$_2$, and CH$_3$OH. 
Whereas for H$_2$CO, the observed value is represented by the horizontal dashed line and shaded region represent a factor of six upper and 
lower abundance compared to the observed value. Each panel show Ca (circle), Cb (triangle), M3 (square), M5 (empty square), and M1 
(x) models for a given species.
}
\label{fig_OBS_COMP}
\end{figure*}

Figure~\ref{fig_OBS_COMP} shows the comparison with the observed range for CO, CO$_2$, CH$_3$OH, and H$_2$CO; and 
Table~\ref{TabOBS} shows observed and peak model abundances in water percentage (first row) as well as relative 
to the total hydrogen (second row) for selected surface species for all the three different physical conditions.
We recall the model descriptions once more. Models were run by varying $\mathcal{R}_{CO}$ 
($E_\textrm{b, CO}$/$E_\textrm{d, CO}$) between 0.1 and 0.5 and designated as M1, M2, M3, M4, and M5, and one model is 
run with coverage-dependent binding energy for CO designated using `C'. Each such model is further classified using the 
alphabet `a' and `b' with $E_\textrm{b, i}$/$E_\textrm{d, i}$ = 0.3 and 0.5 respectively. Abundances are similar 
for M3, M4, and M5 models; therefore these models are represented by the M5 model,
For dense cloud models (Figure~\ref{fig_OBS_COMP}a), 
CO abundances are in the observed range for Cb and M5b models for a considerably large span of time. For models Ca 
and M5a abundances are close to the lower limit of the observed values. Whereas, abundance for the M1 model is 
significantly lower than the observed abundances. For low (Figure~\ref{fig_OBS_COMP}e) and high mass (Figure~\ref{fig_OBS_COMP}i) 
star formation, model abundances fall within the observed range for all the models except model M1a, and M1b. 
It indicates that models for which CO diffusion is relatively slower can explain observed CO abundances better compared to the models
with faster diffusion.

\begin{table*}
\setlength\tabcolsep{2pt}
\centering
\caption {Comparison between observed ice abundances and peak model abundances.}
\label{TabOBS}
\begin{tabular}{|c|l|c|c|c|c|c|c|}
%\begin{tabular*}{\textwidth}{clcccccc}
\hline
Species   & Source          & Observed$^*$   & \multicolumn{5}{|c|}{Models}                          \\
\cline{4-8}
          &                 &               & Ca    & Cb   & M1a - M2a       & M2a - M2b & M5a - M5b \\
\hline
          & Dense           & 9 - 67        & 3     & 30   & 4(-5) - 5(-7)   & 3.5 - 2.8 & 11 - 30   \\
\cline{3-8}
          & clouds (DC)     & 3 - 21        & 1.5   & 25.6 & 1.7(-3) - 2(-5) & 1 - 1.3   & 2.8 - 2.5 \\
\cline{2-8}
CO        & Low-mass        & $<$ 3 - 85    & 8     & 21   & 0.4 - 0.3       & 7.5 - 14  & 8 - 22 \\
\cline{3-8}
          & Protostar (LMP) & $<$ 1.2 - 26  & 10.5  & 28.3  & 0.4 - 0.03     & 10 - 16.2 & 10.8 - 28.7 \\
\cline{2-8}
          & Massive         & 3 - 26        & 8   & 21   & 0.4 - 0.03 & 7.5 - 14 & 8.1 - 22  \\
\cline{3-8}
          & Protostar (MP)  & $<$ 0.4 - 12.8& 10.5  & 28.2 &0.4 - 0.03 &10 - 16.3 & 11 - 28.6\\
\hline
          & Dense           & 14 - 43       & 1.4   & 1.5  & 136 - 144 & 10 - 12 & 1.4 - 2   \\
\cline{3-8}
          & clouds          & 5.2 - 26      & 1.6   & 1.6  & 63.8 - 64.8 & 6.4 - 6 & 1.6 - 2   \\
\cline{2-8}
CO$_2$    & Low-mass        & 12 - 50       & 1.8   & 0.2  &30.8 -35 & 4 - 17.4  & 1.7 - 3.8 \\
\cline{3-8}
          & Protostar       & 2.4 - 25      & 2.4   & 0.31 &31.2 - 33.6 &5.8 - 19.5 & 2.2 - 5 \\
\cline{2-8}
          & Massive         & 11 - 27       & 0.15  & 0.15 & 30.6 - 34.7 & 1 - 12.3 & 0.35 - 0.37  \\
\cline{3-8}
          & Protostar       & 1.8 - 15.6    & 0.19  & 0.19 & 31 - 33.4 & 1.3 - 14.3 &  0.33 - 0.49\\
\hline
          & Dense           & 5 - 12        & 37.5  & 21.8 & 1.8 - 1   & 36 - 4 & 39 - 19.5   \\
\cline{3-8}
          & clouds          & $<$ 0.6 - 6.6 & 37.5  & 27.8 & 0.8 - 0.5 & 35.2 - 2 & 38.5 - 24.5 \\
\cline{2-8}
CH$_3$OH  & Low-mass        & 1 - 30        & 22    & 9    &13.3 - 7.7 & 23 - 8.8 & 23 - 9 \\
\cline{3-8}
          & protostart      & $< 0.2$ - 15  & 28.4  & 9.6  & 13 - 7 &29 - 9 &  29 - 9.7 \\
\cline{2-8}
          & Massive         & 5 - 30        & 22.5  & 9    &12.6 - 0.77 & 23 - 8.8 & 23 - 9 \\
\cline{3-8}
          & Protostar       & $<$ 0.4 - 16.6& 28.4  & 9.6  &12.5 - 7.1 &29 - 9 & 29 - 10 \\
\hline
          & DC              & -             & 3.8   & 16   & 2.3(-2) - 0.9 & 2.9 - 2 & 6.7 - 14  \\
H$_2$CO   & LMP             & $\sim$ 6      & 2     & 4.2  & 0.2 - 4.4 & 2.1 - 4 & 2.1 - 4.2  \\
          & MP              & 1 - 3         & 1.8   & 4.2  & 0.2 -4.4 &1.9 - 4.1 &1.9 - 4.2  \\
\hline
          & DC              & $\sim$ 2      & 6(-5) & 2(-4)& 4(-5) - 4.3(-4) & 6.9(-5) - 4(-4) & 9.3(-5) - 2.4(-4)  \\
HCOOH     & LMP             & 1 - 9         & 0.08  & 0.14  & 0.5 - 0.23  & 0.2 - 0.17 & 0.13 - 0.1 \\
          & MP              & 3 - 7         & 1(-3) & 1(-2)& 3.8(-3) - 1.6(-2) &1(-3) - 2(-3)& 1(-5) - 2(-3)  \\
\hline
          & DC              & $<$ 2         & 2(-8) & 9(-4)& 7.6(-7) - 3.7(-2) & 5(-7) - 3.7(-2) & 3(-7)-3.8(-5)  \\
OCS       & LMP             & -             & 9(-5) & 1(-4)& 6(-6) - 6(-5)& 3(-6) - 4(-5) & 2(-6) - 1(-5)  \\
          & MP              & 0.04 - 0.2    & 5(-5) & 4(-4)& 3(-5) - 1.5(-4) & 3.6(-5) - 1.6(-4) & 2.6(-5) - 1.4(-4)  \\
\hline
\end{tabular}
\\
* Observed abundances are from \cite{Boogert2015} and reference therein. \\
For a given source two sets of abundances are provided: the first row is in water \% and second row is 
relative to n$_H$ ($\times$ 10$^{-6}$) \\
Abundances relative to n$_H$ is given only for CO, CO$_2$ and CH$_3$OH, since for other species abundances are low.
\end{table*}

None of the model results can explain the solid CO$_2$ abundance in cold cores (Figure~\ref{fig_OBS_COMP}b), with an exception 
for the M1a, M1b, and M2b model for the certain time range. Ca, Cb, M3 - M5 models show significantly lower abundance, whereas, 
for model M1a and M1b, peak abundances are nearly three times larger  compared to the upper bound of the observed abundance. Although 
over-produced, the faster diffusion of CO can increase the production of CO$_2$ ice significantly and can match observed abundances 
during the early phase of evolution. For slow and fast warm-up models (Figure~\ref{fig_OBS_COMP}f and j), both the versions 
of M1 and M2b models can match the observed abundance, whereas other models can not produce CO$_2$ in enough quantity to explain 
observed abundances.
In this context it is important to note that to explain the observed abundance of solid CO$_2$, \cite{Garrod2011} incorporated hydrogenation 
of oxygen atom while situated on top of a surface CO molecule thus forming OH on the top of a CO surface. In this prescription, the only 
surface diffusion required to produce CO$_2$ is that of atomic hydrogen not CO or oxygen atom which are slower than atomic hydrogen when 
currently used diffusion barriers are considered. These authors successfully produced observed CO$_2$ abundances using their scheme. It 
implies that the CO diffusion rate should be similar to that of atomic hydrogen to explain observed CO$_2$ abundances.
Besides, higher CO$_2$ abundance could also be found, if the initial dust temperature is 
relatively higher (15 - 20 K). It is found in the star-forming regions of Large and Small Magellanic clouds (\cite{Acharyya2015b, 
Acharyya2016}), where, higher CO$_2$/H$_2$O is observed (dust temperatures are believed to be higher than the dust in our galaxy). Similar 
results were also found by \cite{Kouchi2020}, using the model of \cite{Furuya2015}.
Similarly, \cite{Drozdovskaya2016} found higher solid CO$_2$ abundance in protoplanetary disks via grain surface reaction of OH 
with CO, due to enhanced photodissociation of H$_2$O. Thus the models with low diffusion barrier of solid CO does not improve the model abundance 
of solid CO and CO$_2$ towards the greater agreement with the observed abundances.

The surface abundance of HCO is always very low as can be seen from Figure~\ref{fig_Oth_small}a (cold cores models) and 
Figure~\ref{fig_Oth_small_WU_LM} (slow warm-up models). It is yet to be observed, which is in line with the model results. 
The solid H$_2$CO is yet to be observed in cold clouds, all models over-produces (Figure~\ref{fig_OBS_COMP}c), particularly 
models with slower diffusion (Cb and M5b). It stays within the observed limit for warm-up models for both with the slow and 
fast heating (4th row in Table~\ref{TabOBS} and Figure~\ref{fig_OBS_COMP}g and k) except M1a. For CH$_3$OH, in dense clouds 
(Figure~\ref{fig_OBS_COMP}d), the M1a and M2b models are within the observed abundances. Apart from M1a, M1b and M2b models, 
all the other models tend to over-produce when compared with the observed abundances for most of the time ranges. 
For warm-up models (\ref{fig_OBS_COMP}h and l), abundances for Cb, M1b, M2b models are within observed limits 
(Table~\ref{TabOBS}, 3rd row). 
The abundance of HCOOH for all the cold core models is significantly lower when compared with the observed abundances 
(Table~\ref{TabOBS}, 5th row). The abundance increases for the slow warm-up model but still not sufficient to explain the 
observed abundances. 
Observed solid OCS abundance for cold cores have an upper limit of 2 $\%$ of water, whereas 
it is not observed in low-mass protostars. For massive protostars, its observed range is between 0.04 and 0.2 $\%$ of water 
(Table~\ref{TabOBS}, 6th row). The model abundances are lower than the observed ranges for all the models. 
%Thus, the low diffusion barrier does not have any additional advantage towards explaining the 
%observed abundance.

\section{\rm{Concluding Remarks}} \label{Conclude}
Impact of different diffusion barrier of CO in the grain surface chemistry is studied for three different physical conditions; dense clouds 
and two warm-up models with heating rates, which are representative of low and high mass star formation. For each of these conditions, 
six models were run; one coverage dependent binding energies and five models by varying $\mathcal{R}_{CO}$ 
(E$_\textrm {b, CO}$/E$_\textrm {d, CO}$) between 0.1 and 0.5. Besides for each value of $\mathcal{R}_{CO}$,
two sets of models with $\mathcal{R}_{i}$ (E$_\textrm {b, i}$/E$_\textrm {d, i}$) = 0.3 and 0.5 are run. 
Major conclusions are as follows.

\begin{enumerate}
\item The abundance of CO increases with an increase in $\mathcal{R}_{CO}$, i.e., with decreasing diffusion rate. Abundance 
is higher for $\mathcal{R}_{i}$ = 0.5 compared to 0.3 except M1 models. Models with $\mathcal{R}_{CO} >$ 0.2, and Cb can explain 
the observed abundances, whereas other models, particularly, model M1a, b ($\mathcal{R}_{CO}$ = 0.1) provides significantly 
lower solid CO abundance when compared with the observed abundances due to efficient use of solid CO to form other species owing 
to its faster diffusion. For low and high mass star formation as well, model abundances fall within the observed range except 
for models with $\mathcal{R}_{CO} =0.1$.

\item For solid CO$_2$, none of the models can explain the observed abundances for the dense cloud. The models with faster
diffusion over-produces CO$_2$ by a factor of three. Abundances are within the observed limit for the slow and fast warm-up 
models when $\mathcal{R}_{CO} \le$ 0.2. Formation of CO$_2$ via CO + OH $\rightarrow$ CO$_2$ + H is favoured for 
$\mathcal{R}_{CO} \le$ 0.2 otherwise, CO is hydrogenated to form HCO. 

\item For H$_2$CO and CH$_3$OH agreement with observation can be found for almost all the models albeit in a limited 
range of parameter space. For H$_2$CO, peak abundance comes for $\mathcal{R}_i = $ 0.5 models, whereas for CH$_3$OH, 
it comes for the $\mathcal{R}_i = $ 0.3. For both the species, abundance is lowest for the models with the fastest 
CO-diffusion, since CO is efficiently converted to CO$_2$ instead of getting hydrogenated. 

\item Formation of HCOOH, CH$_3$CHO, NH$_3$CO, and CH$_3$COCH$_3$ depends on the diffusion of CO. Their surface abundances 
differ significantly from one model to the other but, none of the models has any particular edge over others. None of the models 
produces these molecules in enough quantity so that their abundance is above the present-day detection limit on the ice. Among these 
four molecules, only solid HCOOH is likely to be observable, although models with slower diffusion produce 
more solid HCOOH compared to the models with faster diffusion but still not close to the observed abundance.

\item For $\mathcal{R}_{CO} >$ 0.2, the surface abundance of various species involving CO remain almost unchanged. Thus above a 
certain critical diffusion barrier, CO diffusion is slow and can not play a dominant role.

\item  Finally, both $\mathcal{R}_i$ and $\mathcal{R}_{CO}$ plays a crucial role in the formation of molecules, and more 
laboratory measurements are required for both the parameters.

\end{enumerate}

\section*{Acknowledgements}
We thank anonymous referee for constructive comments which strengthened the paper.
The work done at Physical Research Laboratory is supported by the Department of Space, Government of India. 

\newpage
%\section{reference}
\bibliographystyle{pasa-mnras}
\bibliography{export-bibtex}

\end{document}